\documentclass[twocolumn,twocolappendix]{aastex631}

\usepackage[utf8]{inputenc}
\usepackage{newunicodechar} 
\newunicodechar{✱}{\ensuremath{*}}
\usepackage{pifont}
\usepackage{needspace}
\usepackage{amsmath} 

\usepackage{changepage}

\newcommand{\epserib}{$\mathrm{\varepsilon}$ Eridani b}
\newcommand{\epseri}{$\mathrm{\varepsilon}$ Eridani}

\newcommand{\mjup}{\ensuremath{\mathrm{M_{jup}}}}

\newcommand{\nrc}{NRC Herzberg Astronomy and Astrophysics,
5071 West Saanich Road,
Victoria, BC, V9E 2E7, Canada}
\newcommand{\ucsd}{Department of Astronomy \& Astrophysics,  University of California, San Diego, La Jolla, CA 92093, USA}
\newcommand{\nmsu}{Department of Astronomy, New Mexico State University, 1320 Frenger Mall, Las Cruces, NM 88003, USA}
\newcommand{\uvic}{Department of Physics and Astronomy, University of Victoria,
3800 Finnerty Road, Elliot Building,
Victoria, BC, V8P 5C2, Canada}
\newcommand{\ucsc}{
Department of Astronomy \& Astrophysics, University of California, Santa Cruz, CA, USA}

\definecolor{Mycolor1}{HTML}{AA2222}
\definecolor{Mycolor2}{HTML}{0011DE}

\submitjournal{AJ}

\begin{document}

\title{Revised Mass and Orbit of \epserib:\\A 1 Jupiter-Mass Planet on a Near-Circular Orbit}

\author[0000-0001-5684-4593]{William Thompson}
\affiliation{\nrc}

\author[0000-0001-6975-9056]{Eric L. Nielsen}
\affiliation{\nmsu}

\author[0000-0003-2233-4821]{Jean-Baptiste Ruffio}
\affiliation{\ucsd}

\author[0000-0002-3199-2888]{Sarah Blunt}
\affiliation{\ucsc}

\author[0000-0002-4164-4182]{Christian Marois}
\affiliation{\nrc}
\affiliation{\uvic}

\correspondingauthor{William Thompson}
\email{william.thompson@nrc-cnrc.gc.ca}

\begin{abstract}
        The mature Jovian planet \epserib\ orbits one of the closest sun-like stars at a moderate separation of 3.5 AU, presenting one of the best opportunities to image a true analog to a solar system planet.
    We perform a thorough joint reanalysis and cross-validation of all available archival radial velocity and astrometry data, combining data from eight radial velocity instruments
    and four astrometric sources (Hipparcos, Hubble FGS, Gaia DR2, and Gaia DR3).
    We incorporate methodological advances that impact our findings  including a principled treatment of correlation between Gaia DR2 and DR3 velocity.
    We revise the planet's mass upward to $1.00 \pm 0.10 \mjup$ and find that its orbit is nearly circular and likely to be to coplanar with the outer debris disk.
    We further present one of the first models of an exoplanet orbit exclusively from absolute astrometry and independently confirm the planet's orbital period. 
    We make specific predictions for the planet's location at key imaging epochs from past and future observing campaigns.
    We discuss and resolve tensions between previous works regarding the eccentricity, inclination, and mass.
    Our results further support that \epserib\ is one of the closest analogs to a Solar System planet yet detected around a nearby star.

\end{abstract}

\section{Introduction}\label{sec:intro}

\epseri\ is one of the closest planetary systems to Earth, just $3.2$ pc away. 
This proximity has made it a compelling target for direct imaging, giving us an opportunity to potentially image a true analogue to a planet in our Solar System: a relatively mature (200-800 Myr old, \citet{mamajek_2008}; 439 or 659 Myr old, \citet{Barnes_2007}) 
near-Jupiter mass planet, orbiting a Sun-like star at a separation that would fall between Mars and Jupiter ($\approx3.5$ AU).

This exosystem has been extensively studied since the 1980s, when pioneering Doppler radial velocity (RV) surveys by \citet{rvs_cfht_campbell_1988} detected a long-term trend, and then \citet{epseri_lick_Hatzes_2000} confirmed periodic motion of the star.
Further credence was given to these RV signals when \citet{fgs_eps_eri_benedict_2006} determined a complete orbit by combining RV data with astrometry from the Hubble Telescope's Fine Guidance Sensor (FGS), the Allegheny Multichannel Astrometric Photometer (MAP)
\citep{map_astrometry_gatewood_1987,epseri_proceeding_gatewoord_2000}, and Hipparcos \citep{Hipparcos_van_leeuwen_1997}, making it one of the earliest planets with an orbit derived in-part from absolute astrometry.

Since this time, studies of \epserib\ have entered a renaissance, driven by the promise that high-contrast direct imaging technology has advanced to the point where we may, in the near future, be able to study the planet directly via its thermally emitted light.

Several major attempts to image this system in the near-infrared have been published to date, including 
K-band imaging at Keck by \citet{epseri_kband_macintosh_2003},
$\mathrm{L^\prime}$ and M-band imaging with MMT by \citet{epseri_imaging_mmt_heinze_2008},
L-band imaging with VLT by \citet{epseri_janson_2008},
L- and M-band imaging with Spitzer by \citet{epseri_janson_2015},
M-band imaging with Keck by \citet{epseri_mawet_2019},
M-band imaging with Keck by \citet{epseri_jorge_2021},
and N-band at VLT imaging with \citet{epseri_near_pathak_2021}.
Though these efforts did not achieve an imaging detection, they placed upper limits of  on the brightness of the planet, and therefore limit its mass to  $\lesssim2 \,\mjup$ for a 400 Myr system age.

Motivated by these attempts, upcoming and completed observations from JWST using NIRCam, MIRI, and NIRSpec are poised to reach significantly better sensitivity. This renews the need to make specific predictions of the planet's position in order to 
determine the best epochs to observe the planet,
 measure its mass and thus expected brightness,
and confirm the location of any observed point sources.
In addition, certain observations could benefit from knowing the location of the planet in advance to adjust the detector placement to mitigate saturation, plan dither patterns, and so on.

The most recent works to re-examine the orbit of \epserib\ include \citet{HipparcosIAD_planets_Reffert_2011}, \citet{eps_eri_chiron_Giguere_2016}, \citet{epseri_jorge_2021}, \citet{epseri_express_Roettenbacher_2022}, \citet{fgs_eps_eri_benedict_2020}, and \citet{epseri_epsindi_feng_2023}. 
These works examined the orbit of \epserib\ using a mix of data from various RV sources, Hipparcos \citep{hipparcos_van_Leeuwen_2007},   Gaia DR2 \citep{gaia_dr2}, and Gaia DR3 \citep{gaia_dr3}, but arrived at somewhat inconsistent conclusions with regard to the planet's orbital eccentricity, inclination, and dynamical mass.

The goal of this paper is to determine the orbit of \epserib\ using a careful reanalysis of all available archival data.
We combine the superset of all observations considered in previous works, including radial velocity data from eight RV instruments (APF, HIRES, HARPS, Lick, CFHT, CHIRON, EXPRES, and CES) and four astrometric data sources (Hipparcos, Hubble FGS, Gaia DR2, and DR3).
Throughout, we take special care to account for systematics and second-order effects that become non-negligible for fast-moving stars over long time-scales, including rigorous propagation of the system's barycentric motion, perspective acceleration, changing parallax, and changing light travel-time to the receding system.   
We compare and contrast the methods and results of several previous analyses, finding where they agree and identifying which assumptions had led to disagreements. 

In anticipation of a successful imaging detection, \epserib\ can also serve as a test run for future deep imaging campaigns with JWST, Roman, or HWO.
Some of these campaigns will depend on radial velocity and astrometry to determine the location and brightness of planets ahead of time. Due to the long integrations required to detect sub-Jovian planets, it is incumbent on the community to demonstrate the ability to robustly determine planet positions and brightness ahead of targeted observations.
We hope that a correct prediction of the location of \epserib\ can demonstrate this capability.

We now begin this paper by describing the sources of data we use, how that data is treated in our models, and other details of our modeling approach. We then present results, broken down by different subsets of data. Finally, we discuss how our results compare to previous efforts and the implications of our work.

\section{Data and Methods}

\subsection{Radial Velocity Data}

We gathered radial velocity data from many sources, starting with some of the earliest RV measurements made in the 1980s by CFHT \citep{rvs_cfht_campbell_1988}. We used the ``corrected'' data presented in Table 1. 
We used data from Lick gathered from \citet{rvs_lick_fisher_2014}, and split it into four groups as noted by the authors:
Lick (1), before November 1994;
Lick (2), dewar 13;
Lick (3), dewar 6;
Lick (4), dewar 8.
We used data from CES \citep{rvs_ces_Zechmeister_2013}, both from the LC and VLC, and treated these data separately.
We further included data from HARPS \citep{rvs_harps_Trifonov_2020}, split between before and the 2015 upgrade (``pre'' and ``post'' respectively).
We used all HIRES and APF RVs presented by \citet{epseri_mawet_2019} and by \citet{epseri_jorge_2021}. 
We further added RVs from CHIRON reported by \citet{eps_eri_chiron_Giguere_2016} and from  EXPRES reported by \citet{epseri_express_Roettenbacher_2022}. 

This gives nearly complete coverage of the orbit for 40 years, between 1981 and 2021, in thirteen separate groups of RV data.

Over such a long time-scale, 3D perspective effects caused by the star's high proper motion ($\approx$ 1"/yr) and barycentric radial velocity (16 km/s) have the potential to build up to detectable levels.
We elected to handle this apparent acceleration term directly in our models and thus attempted to undo any pre-existing static corrections applied to each dataset.
We assumed that this correction was not applied for the CFHT, CES, HIRES, CHIRON, and EXPRES datasets 
and assumed that the correction was applied for the HIRES, APF, HARPS, and Lick datasets.
For the latter datasets, we added back in secular acceleration at the rate given by Gaia DR3.
To be conservative, we nonetheless include a linear trend for each dataset in our model. 
We encourage authors to include descriptions of these corrections in their works so that future authors can make the best use of their data, especially for systems with less orbital coverage than \epseri.

To model the RV data, we used a separate offset ($\gamma$), jitter ($\sigma$), and slope parameters ($m$) for each group, adding 39 variables to our model.
We modeled the activity signal from the star, which has been well established as having an $\approx 11 \mathrm{d}$ signal from rotational modulation of the star and having an approximately month-long decay timescale.
We use the Celerite \citep{celerite_dfm} framework through the \verb|Celerite.jl| package\footnote{\href{https://github.com/ericagol/celerite.jl/}{https://github.com/ericagol/celerite.jl/}} and define an approximate quasi-periodic GP kernel with parameters $B$, $C$, $L$, and $P_\mathrm{rot}$, just as in \citet{epseri_express_Roettenbacher_2022}. 

We modeled each dataset with an independent Gaussian process, and multiplied their likelihoods. This treatment inherently assumes that interleaved data from different instruments are uncorrelated (see \citet{Blunt_2023}, section 4). As discussed in Section \ref{sec:discussion-ecc}, we tested the impacts of this assumption by comparing against a model with a single GP shared across all instruments, ultimately finding that our results in this case are not sensitive to the exact treatment.

\subsection{Astrometric Data}

The \epseri\ system has been monitored for astrometric motion and parallax for over a century.
We include precision absolute astrometry from Hipparcos \citep{hipparcos_van_Leeuwen_2007}, Hubble Fine Guidance Sensor \citep[FGS, from ][]{fgs_eps_eri_benedict_2006}, and both Gaia's second \citep[DR2, ][]{gaia_dr2} and third data releases \citep[DR3, ][]{gaia_dr3}.

Beginning with Hipparcos, we used data from the mission in two different ways.
We used the Hipparcos IAD epoch astrometry sourced from the \emph{Java Tool} following the guidance of \citet{betapic_nielsen_2020}, renormalizing the uncertainties, and testing that we could reproduce the reported catalog measurements and uncertainties exactly.
These data were used for the Hipparcos IAD model to be described shortly.
We also used the Hipparcos proper motion sourced from the DR3 Hipparcos-Gaia catalog of accelerations \citep[HGCA;][]{hgca_brandt_2021}, calculated from a mix of the 1997 and 2007 Hipparcos data releases, and calibrated against the Gaia DR3 velocity reference frame. 
These data were used in all other relevant astrometry models to ensure the reference frame was well-calibrated against that of Gaia DR3. 
This is a key departure from \citet{epseri_epsindi_feng_2023}, who did not use the HGCA calibration between Gaia and Hipparcos reference frames, electing to include the reference frame alignment as variables in their model.

The second source of absolute astrometry data we used for this system was that of \citet{fgs_eps_eri_benedict_2006}. 
The data used in that publication were derived by measuring the positions of \epseri\ using the Hubble Fine Guidance Sensor (FGS) relative to a field of nearby background stars.
Although the raw data are available, the re-reduction of these data is out of scope for this work. Instead, we opted to extract the residuals from their parallax and proper motion fit (Figure 7 of \citet{fgs_eps_eri_benedict_2006}) using \emph{Web Plot Digitizer} \citep{WebPlotDigitizer}.

The extracted residuals are relative to a five-parameter linear model of annual parallax and proper motion fit to a combination of the FGS data, lower-precision astrometry from the Allegheny Multichannel Astrometric Photometer (not considered in this work).
We justify using these residuals in our model by noting that the authors' best-fit five-parameter model arrived at a parallax value that is very close to the measurement from Gaia DR3, meaning that no annual cyclic motion should be detectable in the residuals.
Then, any residual proper motion not captured by their best-fit five-parameter model would appear as a linear trend.
Such a linear trend could also be caused by frame rotation versus the Gaia DR3 reference frame, which we must assume is present to some degree.
Rather than attempt to calibrate the absolute astrometry against Gaia with sufficient accuracy, we elect to simply fit linear trends to the residuals and fit only the acceleration over the FGS observations, which will be the same in any inertial reference frame. This is analogous to including a linear trend in models of RV data. This neatly sidesteps the issue of determining an absolute calibration of the FGS data and allowing us to use the residuals directly.

The final sources of absolute astrometry are both from Gaia.
We include Gaia DR3 proper motion and uncertainties from the HGCA, just as we do the Hipparcos and Gaia-Hipparcos long-term proper motion.
We undo the non-linear corrections incorporated by the HGCA to account for perspective acceleration and curvature in the R.A. and Dec. coordinate system, in favor of handling this motion inside our model.

We further include the Gaia DR2 catalog proper motion in some models. 
The DR2 and DR3 velocities can provide acceleration constraints because, despite sharing the majority of data between the two catalogs, the DR2  catalog values covers a time baseline that is $\approx 66\%$ shorter.
Thus, despite the overlapping source data, a change in the velocity reported by the two catalogs can still provide some modest constraints on the acceleration of the star.

We model the two catalogs by simulating Gaia observations at each visibility window reported by GOST \footnote{\href{https://gaia.esac.esa.int/gost/index.jsp}{https://gaia.esac.esa.int/gost/index.jsp}}. We remove known gaps in Gaia data, leaving only one extra epoch from GOST unaccounted for. We include this visibility window as a discrete parameter in our model and marginalize over it.

We then perform a linear least-squares fit of a 5-parameter model to both the DR2 time-range and the DR3 time-range.
We then compared the velocities extracted from these simulations, $\mu_\mathrm{sim} = [\mu_\mathrm{DR2,sim} \; \mu_\mathrm{DR3,sim}]$ against the reported catalog values $\mu = [\mu_\mathrm{DR2} \; \mu_\mathrm{DR3}]$.

This is similar in principle to the work of \citep{epseri_epsindi_feng_2023}, though we differ in how the data are modeled a few important respects. First, we handle the correlation between DR2 and DR3, and second, we use the HGCA's spatially dependent calibration between the Hipparcos and Gaia reference frames and calibrated (inflated) uncertainties rather than those reported by Gaia directly combined with flexible jitter parameters.

The reason we must model the correlation between DR2 and DR3 is that the reported velocities are derived from as much as $\approx60 - 70\%$ of the same measurements.
We therefore form a full covariance matrix between R.A. and Dec. velocities from both catalogs as the block matrix:
\begin{equation}
    \Sigma_{DR2,DR3} = \begin{bmatrix}
        \Sigma_\mathrm{DR2}      & \Sigma_\mathrm{DR2,DR3} \\
        \Sigma_\mathrm{DR2,DR3}  & \Sigma_\mathrm{DR3}
    \end{bmatrix}
\end{equation}
While the DR2 and DR3 catalogues report the covariance between derived R.A. and Dec. velocities, $\Sigma_\mathrm{DR2}$ and $\Sigma_\mathrm{DR3}$ respectively, we must determine an appropriate form for the inter-cataloge covariance $\Sigma_\mathrm{DR2,DR3}$.

We estimate the off-diagonal blocks $\Sigma_\mathrm{DR2,DR3}$ of the covariance matrix using the reported uncertainties from Gaia DR2 and DR3 and the fraction of shared epochs. We assume that the sampling is uniform in time. Under these assumptions, consider a 1-D case with uniform time sampling, and the calculation of slope $\mu_A$ of a set of $M$ independent Gaussian random variables with uniform uncertainties, and the slope $\mu_B$ of a set of N independent Gaussian random variables which include the same set of $M$ (where $N>M$) of random variables as in $\mu_A$ (just as Gaia DR3 includes all measurements from Gaia DR2, plus an addition set of new measurements).

To determine the correlation between $\mu_A$ and $\mu_B$, we start with the least squares slope of a line which is given by 
$$
\mu = \frac{\sum_{i=1}^n (x_i - \bar{x})(y_i - \bar{y})}{\sum_{i=1}^n (x_i - \bar{x})^2}
$$
This can be rearranged to show that the slope is a linear combination of the y-values:
$$
\mu = \sum_{i=1}^n w_i y_i
$$
where 
$$w_i = \frac{x_i - \bar{x}}{\sum_{j=1}^n (x_j - \bar{x})^2}$$
Therefore, our two slopes can be written as:
$$
\mu_A = \sum_{i=1}^{M} w_i X_i
$$
and 
$$
\mu_B = \sum_{i=1}^{M} w_iX_i + \sum_{i=M+1}^{N} w_{i}X_i
$$
where $X_i$ are independent normal random variables with variance $\sigma^2$.
The variances are:

$$
Var(\mu_A) = \sum_{i=1}^{M} w_i^2\sigma^2
$$
and
$$
Var(\mu_B) = \sum_{i=1}^{M} w_i^2\sigma^2 + \sum_{i=M+1}^{N} w_{i}^2\sigma^2,
$$
and the covariance is
$$
Cov(\mu_A,\mu_B) = \sum_{i=1}^{M} w_i^2\sigma^2.
$$
Giving, therefore, the corelation
$$
\rho = \frac{Cov(\mu_A,\mu_B)}{\sqrt{Var(m_A)Var(m_B)}}
$$
$$
= \frac{\sum_{i=1}^{M} w_i^2\sigma^2}{\sqrt{\sum_{i=1}^{M} w_i^2\sigma^2 \cdot (\sum_{i=1}^{M} w_i^2\sigma^2 + \sum_{i=M+1}^{N} w_{i}^2\sigma^2)}}
$$

$$
= \frac{M}{{\sqrt{M \cdot N}}} = \sqrt{\frac{M}{N}}
$$

This demonstrates that the correlation between the two slopes depends only on the proportion of shared variables, under these idealized assumptions.
We therefore treat the Gaia DR2 proper motion in either R.A. or Dec. as having the correlation $\rho$ with the proper motions reported by DR3.

The last detail to complete is the cross terms giving covariance between the DR2 proper motion in R.A. and DR3 proper motion in Dec., and vice-versa. 
 
We simply generalize the covariance matrix between two scalar variables,
$$
\Sigma_{a,b} = \begin{bmatrix}
    \sigma_a^2           &   \rho\sigma_a\sigma_b \\
    \rho\sigma_a\sigma_b &   \sigma_b^2
\end{bmatrix}
$$ to the matrix case
$$
\Sigma_{DR2,DR3} = \begin{bmatrix}
    \Sigma_\mathrm{DR2}          & \rho\sqrt{\Sigma_\mathrm{DR2}}\sqrt{\Sigma_\mathrm{DR3}} \\
    \rho\sqrt{\Sigma_\mathrm{DR2}}\sqrt{\Sigma_\mathrm{DR3}}  & \Sigma_\mathrm{DR3}
\end{bmatrix}
$$ where $\rho=\sqrt{\frac{M}{N}}$ as above and where $\sqrt{\Sigma_\mathrm{DR2}}$ and  $\sqrt{\Sigma_\mathrm{DR3}}$ are the matrix square roots, which are unique for positive-definite matrices like covariance matrices.

One additional consideration is that the DR2 proper motion reference frame is known to be rotating slightly compared to DR3. We corrected for this according to
$(\omega_X, \omega_Y, \omega_Z) = (-0.068 \pm 0.051, -0.051 \pm 0.042, -0.014 \pm 0.036)$ from \citet{dr2_reference_frame_Lindegren_2020}. 
Unlike previous works that simply adopted the mean values of these parameters, we included them as priors in our models in order to rigorously marginalize over the uncertainties in the reference frame calibration.
This is similar in effect to the offsets employed by \citet{epseri_epsindi_feng_2023}, but only applied to the DR2-DR3 velocity difference.

As noted by the Gaia teams and in the HGCA \citep{gaia,hgca_brandt_2018,hgca_brandt_2021}, the formal uncertainties reported by Gaia and Hipparcos are underestimated. Accordingly, the the HGCA catalogs inflate the proper motion uncertainties such that a sample stars that are not likely to be accelerating have standardized residuals that follow a Gaussian distribution. This error inflation factor for Gaia DR2 and DR3 varies with spatial position.
Where \citet{epseri_epsindi_feng_2023} adopts a flexible jitter parameter, we used the calibrated uncertainties from the HGCA catalogs for both DR2 and DR3.

Based on all these considerations, the covariance matrix we adopt  between Gaia DR2 and DR3 R.A. and Dec. proper motions is
$$
\Sigma_{DR2,DR3} = \begin{bmatrix}
  0.752 &   0.154 &  0.156   & -0.007\\
  0.154 &   0.723 & -0.002 &  0.114\\
  0.156 &  -0.007 &  0.048  & -0.010\\
 -0.002 &   0.114 & -0.010  &  0.027\\
\end{bmatrix}\;,
$$
which corresponds to the correlation matrix
$$
C_{DR2,DR3} = \begin{bmatrix}
  1.0    &   0.208 &  0.821  & -0.017 \\
  0.208  &   1.0   & -0.037  &  0.813 \\
  0.821  &  -0.037 &  1.0    & -0.278 \\
 -0.017  &   0.813 & -0.278  &  1.0   \\
\end{bmatrix}\;.
$$
We note the high correlations, as expected, between DR2 and DR3 proper motions along the same axes, moderate correlation between the proper motions along different axes in the same data release, and low correlation between the axes of the different data releases.

\subsection{Imaging}
There is raw image data available from previous campaigns, including those by \citet{epseri_mawet_2019}, \citet{epseri_jorge_2021}, and \citet{epseri_near_pathak_2021}.
We did not include these images in our model since the published sensitivity curves indicate that the planet would be below the level of the noise for reasonable inclination angles (limit of $\approx2.2 \,\mjup$ from \citet{epseri_mawet_2019}, 2–4 \mjup\ from \citet{epseri_near_pathak_2021}).
A future work could explore integrating image data into this orbit model following the same recipes as \citet{epseri_jorge_2021} and \citet{hr8799_Thompson_2023}, especially if more images of comparable or greater sensitivity become available.

\subsection{Non-Linear Effects}
In order to rigorously account for second order acceleration terms originating from the system's close proximity to Earth, we model the three-dimensional motion of the system's barycenter through space.

We convert the  proper motion, parallax, and radial velocity at a reference epoch, taken to be J2016, 
into Cartesian coordinates and propagate them linearly versus time assuming uniform space motion.
This accounts for perspective (secular) acceleration, for curvature in the R.A. and Dec. spherical coordinate system, and for changing parallax versus time which causes an slight apparent magnification versus time effect.

In addition, we include the changing light travel-time to the system in our model.
RV and pulsar timing data are often corrected for the so-called ``Rømer-delay,'' the changing light travel time between a source and the Earth due to the Earth's travel around the Sun.
In addition to the Earth's motion, we note that the distance from our solar system to \epseri\  has increased by 19 light-hours between 1980 and 2020. 
This has a small effect on the apparent orbital motion of the planet on the order of 1 mas, and on the perturbed stellar radial velocity on the order of 2 cm/s. It also has an impact of the observed long-term proper motion on the order of 2 mas less displacement than expected from straight-line motion extrapolated over this time span.

To account for this changing light travel time in our models, we solve for the emission time $t_{em}$ at each epoch that corresponds to our observation time $t_{obs}$ using an iterative scheme.
We calculate this difference compared to the reference epoch J2016, for which we set $t_{em} = t_{obs}$; that is, we don't consider the total light travel time, only the change in light travel time from the system's distance at the reference epoch.

We make these corrections at every epoch and model evaluation since the barycentric proper motion of the system is allowed to vary, and projecting the (near-) instantaneous proper motion from Gaia back would unintentionally mix in small contributions from the planet's orbit.

We do not consider the third order perturbations to this motion caused by the planet; that is, the above effects are calculated for the system's barycenter applied to both the star and planet.

Although we include the light travel time effect in our model, for typical stars and timescales it is very well approximated as a linear function appearing to increase the proper motion measured by both Hipparcos and Gaia. As such, readers should note that our proper motion parameters do not include this term and are therefore approximately 0.05 mas/yr lower than the apparent sky motion.
Future orbit models of precision data, e.g. combining Gaia DR4 and high precision interferometry should take this effect into account when modeling long-term proper motion.

\subsection{Gaussian Process}

\epseri\ has significant photometric and RV variability consistent with a rotation period of 10-12 days.
As with the works of \citet{epseri_jorge_2021}, \citet{eps_eri_chiron_Giguere_2016}, and \citet{epseri_express_Roettenbacher_2022}, we model this activity signal using a Gaussian process with a quasi-periodic kernel. 
We implemented the GP in our model using the Celerite.jl package from \citet{celerite_dfm}, after forking and updating the package to work in modern Julia.
We fit the GP separately to each dataset, but share a single set of GP hyper-parameters between all instruments.

\needspace{6em}
\subsection{Implementation}
We implemented our models in the Julia-based \citep{bezanson_julia_2012} Octofitter framework \citep{thompson_octofitter_2023} and sample using stabilized variational non-reversible parallel tempering \cite{stabvarpartemp,nonreversept}
implemented in Pigeons.jl \citep{pigeons}.

Unlike the model presented in \citet{epseri_jorge_2021}, and like the models presented in \citet{epseri_epsindi_feng_2023} and \citet{fgs_eps_eri_benedict_2020} we jointly model both the astrometry and radial velocity, rather than model the RV and astrometry separately.
Unlike \citet{epseri_epsindi_feng_2023} and \citet{fgs_eps_eri_benedict_2020}, we include the Gaussian process model of stellar activity. 
As with \citet{fgs_eps_eri_benedict_2020} we include the Hubble FGS astrometry originally published in \citet{fgs_eps_eri_benedict_2006}.

The most complete model which incorporates every listed data source has a total of 66 different parameters. The majority of these are instrument-specific nuisance parameters, which we marginalize over when presenting results.
The full list of parameters and our adopted priors are listed in Table \ref{tab:variables}.
Whenever we sample from a model that ignores a certain dataset, we remove the related instrument-specific nuisance parameters (e.g. a model without any RV data does not include the many RV offsets, jitters, and slope parameters).

We sample each model until the log-evidence converges between sampling rounds and the $\widehat{\mathrm{R}}$ diagnostic \citep{rhat_gelman,rhat_vehtari_2021} is $\approx 1$ to at least two or more decimal places.

\startlongtable
\begin{deluxetable*}{llcl}
\tablecaption{
List of parameters, descriptions, units, and \textbf{adopted priors}.\label{tab:variables}}
\tablehead{\colhead{Parameter} & \colhead{Description} & \colhead{Unit} & \colhead{Prior}}
\startdata
$m_{A}$ & Primary mass & $\mathrm{M\_\odot}$& $ 0.82 \pm 0.02 $     \\
$m_{b}$ & Secondary mass & $\mathrm{M\_jup}$& Uniform(0, 10)     \\
$P$ & Orbital period  & yr&  Uniform(5.36, 9.36)   \\
$e$ & Orbital eccentricity & &  Uniform(0, 0.999)    \\
$i$ & Orbital inclination & deg&  Sine      \\
$\Omega$ & Longitude of ascending node & deg& Uniform(0,360)     \\
$\tau$ & Orbit fraction at reference epoch &  &   Uniform(0,1)   \\
$\omega$ & Argument of periastron & deg&  Uniform(0,360)     \\
$B$ & GP amplitude & m/s&   Uniform(0.00001, 2000000)   \\
$C$ & GP amplitude ratio & &  Uniform(0.00001, 200)    \\
$L$ & GP decay timescale & days&  Uniform(2, 200)    \\
$P_{rot}$ & GP rotation period & days&  Uniform(8.5, 20)    \\
$\gamma_{APF}$ & APF zero point & m/s&  Uniform(-1000, 1000)   \\
$\gamma_{EXPRES}$ & EXPRES zero point & m/s&  Uniform(-1000, 1000)    \\
$\gamma_{CHIRON}$ & CHIRON zero point & m/s&  Uniform(-1000, 1000)    \\
$\gamma_{CFHT}$ & CFHT zero point & m/s&   Uniform(-1000, 1000)   \\
$\gamma_{HIRES}$ & HIRES zero point & m/s& Uniform(-1000, 1000)    \\
$\gamma_{Lick_1}$ & Lick (1) zero point & m/s&   Uniform(-1000, 1000)  \\
$\gamma_{Lick_2}$ & Lick (2) zero point & m/s&   Uniform(-1000, 1000)   \\
$\gamma_{Lick_3}$ & Lick (3) zero point & m/s&   Uniform(-1000, 1000)   \\
$\gamma_{Lick_4}$ & Lick (4) zero point & m/s&  Uniform(-1000, 1000)    \\
$\gamma_{HARPS,pre}$ & HARPS (pre) zero point & m/s&  Uniform(-1000, 1000)    \\
$\gamma_{HARPS,post}$ & HARPS (post) zero point & m/s&  Uniform(-1000, 1000)    \\
$\gamma_{CESlc}$ & CES (lc) zero point & m/s&  Uniform(-1000, 1000)   \\
$\gamma_{CESvlc}$ & CES (vlc) zero point & m/s&  Uniform(-1000, 1000)    \\
$\sigma_{APF}$ & APF jitter & m/s&  Uniform(0.1, 100)    \\
$\sigma_{EXPRES}$ & EXPRES jitter & m/s& Uniform(0.1, 100)     \\
$\sigma_{CHIRON}$ & CHIRON jitter & m/s& Uniform(0.1, 100)     \\
$\sigma_{CFHT}$ & CFHT jitter & m/s&   Uniform(0.1, 100)   \\
$\sigma_{HIRES}$ & HIRES jitter & m/s& Uniform(0.1, 100)     \\
$\sigma_{Lick_1}$ & Lick (1) jitter & m/s& Uniform(0.1, 100)     \\
$\sigma_{Lick_2}$ & Lick (2) jitter & m/s&  Uniform(0.1, 100)    \\
$\sigma_{Lick_3}$ & Lick (3) jitter & m/s&  Uniform(0.1, 100)    \\
$\sigma_{Lick_4}$ & Lick (4) jitter & m/s&  Uniform(0.1, 100)    \\
$\sigma_{HARPS}$ & HARPS jitter & m/s&  Uniform(0.1, 100)    \\
$\sigma_{CESlc}$ & CES (lc) jitter & m/s&  Uniform(0.1, 100)    \\
$\sigma_{CESvlc}$ & CES (vlc) jitter & m/s&  Uniform(0.1, 100)    \\
$m_{APF}$ & APF slope & m/s/day&  $ 0 \pm 10 $    \\
$m_{EXPRES}$ & EXPRES slope & m/s/day& $ 0 \pm 10 $     \\
$m_{CHIRON}$ & CHIRON slope & m/s/day& $ 0 \pm 10 $     \\
$m_{CFHT}$ & CFHT slope & m/s/day&   $ 0 \pm 10 $   \\
$m_{HIRES}$ & HIRES slope & m/s/day&  $ 0 \pm 10 $    \\
$m_{Lick_1}$ & Lick (1) slope & m/s/day&  $ 0 \pm 10 $    \\
$m_{Lick_2}$ & Lick (2) slope & m/s/day&  $ 0 \pm 10 $    \\
$m_{Lick_3}$ & Lick (3) slope & m/s/day&  $ 0 \pm 10 $    \\
$m_{Lick_4}$ & Lick (4) slope & m/s/day&  $ 0 \pm 10 $    \\
$m_{HARPS,pre}$ & HARPS (pre) slope & m/s/day&  $ 0 \pm 10 $    \\
$m_{HARPS,post}$ & HARPS (post) slope & m/s/day&  $ 0 \pm 10 $    \\
$m_{CESlc}$ & CES (lc) slope & m/s/day&   $ 0 \pm 10 $   \\
$m_{CESvlc}$ & CES (vlc) slope & m/s/day& $ 0 \pm 10 $     \\
$\omega_{x}$ & DR2 velocity offset, x & mas/yr& Normal(-0.068, 0.051)     \\
$\omega_{y}$ & DR2 velocity offset, y & mas/yr& Normal(-0.051, 0.042)     \\
$\omega_{z}$ & DR2 velocity offset, z & mas/yr& Normal(-0.014, 0.036)     \\
$\varpi$ & Parallax & mas&  Uniform(310.577, 0.135)  \\
$\mu_{\alpha✱}$ & Proper motion in RA & mas/yr& $975 \pm 25  $     \\
$\mu_{\delta}$ & Proper motion in Dec & mas/yr& $20 \pm 25$    \\
$\sigma_{FGS}$ & Fine Guidance Sensor jitter  & mas&   LogUniform(0.1, 100)   \\
$\gamma_{FGS\_dec}$ & FGS offset in $\delta$ & mas & Uniform(-500,500) \\
$\gamma_{FGS\_ra}$ & FGS offset in $\alpha✱$ & mas & Uniform(-500,500) \\
$m_{FGS\_\alpha✱}$ & FGS trend in $\alpha✱$ & mas/d 7 & Uniform(-10,10)\\
$m_{FGS\_\delta}$ & FGS trend in $\delta$ & mas/d 7 & Uniform(-10,10)\\
\enddata
\end{deluxetable*}

\ 
\needspace{12em}
\subsection{List of Models}

In order to assess the agreement between different datasets and rule out systematic errors, we create a set of 7 different models. We progressively add additional data to build up to the complete model.
We compare the parameters inferred by each model and their predictions for the planet's location to study what constraints come from which dataset.
Checking that the inferred parameters and predictions are not in tension between models will allow us to check for signs of systematic errors that could bias our results.

The set of seven main models we consider are:
\begin{enumerate}
    \item All RV data  (no astrometry)
    \item All astrometry data (no RV)
    \item RV and Hubble FGS astrometry
    \item RV and Hipparcos IAD (capturing just the motion within the Hipparcos mission)
    \item RV and Hipparcos-Gaia acceleration from the HGCA \citep{hgca_brandt_2021} (capturing the net motion at the Hipparcos and Gaia epochs, versus their long-term trend between missions)
    \item RV and HGCA, with the addition of the Hubble FGS astrometry 
    \item Full model with RV, HGCA, FGS, and the DR2 vs. DR3 acceleration.
\end{enumerate}

The 1\textsuperscript{st} (RV only) and 2\textsuperscript{nd} (astrometry only) models include entirely separate data, aside from the position and parallax of the star from Gaia DR3. 
Similarly, if we consider that the orbital semi-major axis, eccentricity, and argument of periastron are determined largely by RV data, we can then also compare the nearly independent constraints on the inclination and position angle of ascending node from the Hipparcos IAD, the Gaia-Hipparcos acceleration, and the Hubble FGS data using results from the 3\textsuperscript{rd}, 4\textsuperscript{th}, and 5\textsuperscript{th} models.
Note that the Hipparcos IAD residuals are only used in the 3rd model---the others use the calibrated Hipparcos catalog parameters from the HGCA and only use the measurement epochs from the IAD.

\begin{figure*}
    \begin{adjustwidth}{-1.45cm}{-1.2cm} 
    \includegraphics[width=\linewidth]{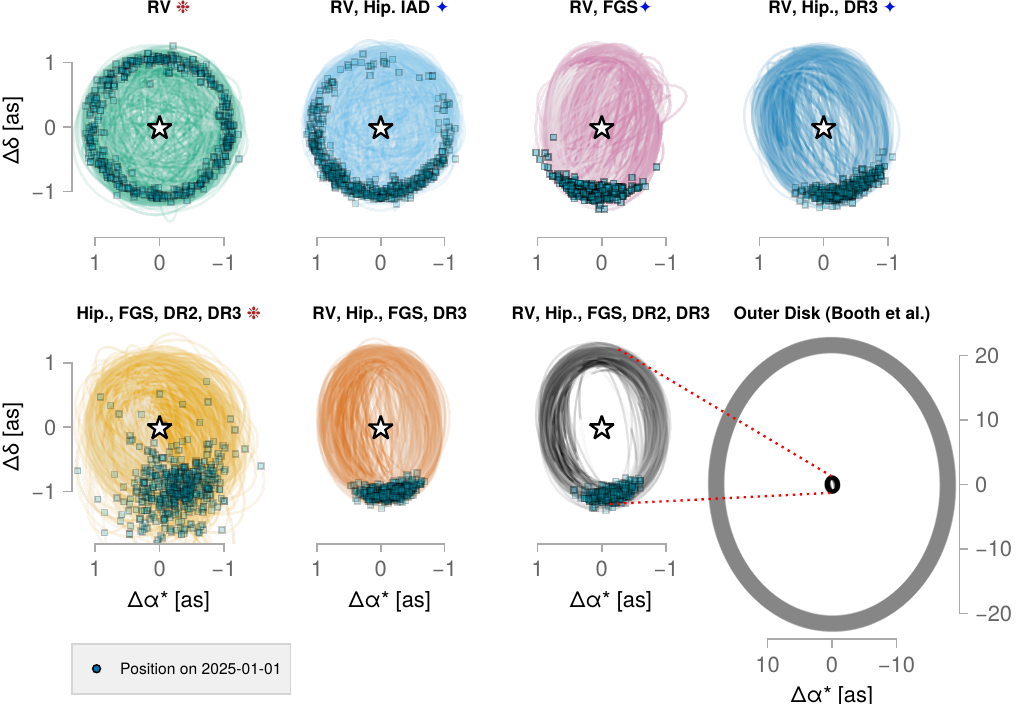}
    \end{adjustwidth}
    \caption{Comparison of the orbit posteriors incorporating different sources of data. The blue dots mark the location of the planet on 2025-01-01.
    The colors of the orbit draws match those in Figure \ref{fig:model-comparison} for easier cross-referencing.
    The bottom right panel shows the outer debris disk model fit to ALMA data by \citet{epseri_alma_booth_2023}.
    The \textcolor{Mycolor1}{\ding{105}} indicates two models containing only RV data and only absolute astrometry data respectively, and whose posteriors are therefore completely independent.
    The models indicated by \textcolor{Mycolor2}{\ding{70}} include RV data, but distinct sets of astrometric data, making the position angles they indicate on 2025-01-01 effectively independent.
    \label{fig:orbit-comparison}}
\end{figure*}

\begin{figure*}
    \centering
    \vspace{-2em}
    \includegraphics[width=\linewidth]{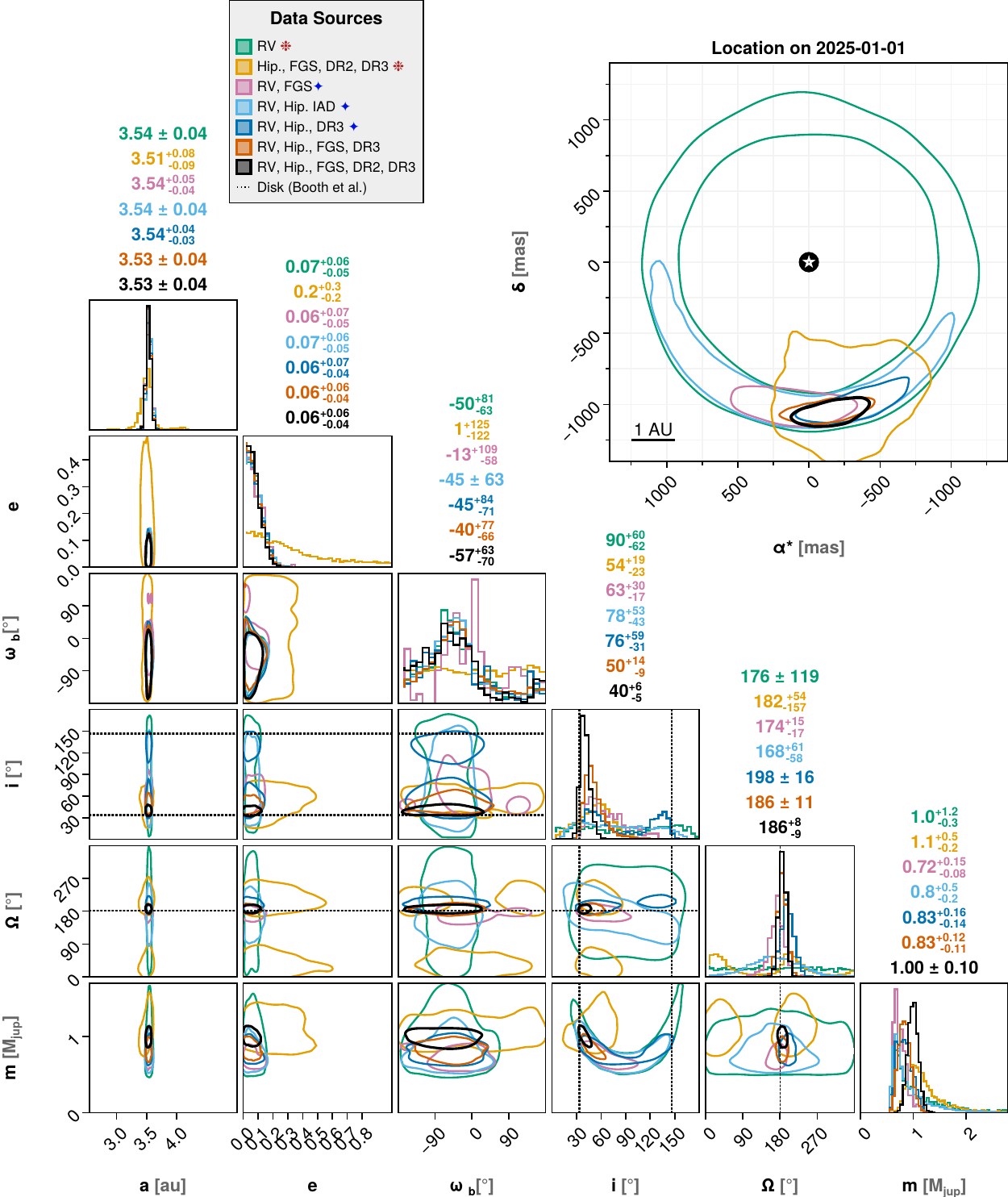}
    \caption{
        A comparison between posteriors from seven models orbit models, listed in the legend. We find no obvious discrepancy between any dataset.
        The bottom left panel is a corner plot, and the top right panel shows the predicted location on 2025-01-01, near the planets maximum separation from the star and the epoch of recent JWST observations. All contours encompass the volume contained within the $1\sigma$ 2D Gaussian equivalent.
        The inclination and position angle of the outer disk model from \citet{epseri_alma_booth_2023} fitted to ALMA data is over-plotted as dashed lines.
        The \textcolor{Mycolor1}{\ding{105}} indicates two models containing only RV data and only absolute astrometry data respectively, and whose posteriors are therefore completely independent.
        The models indicated by \textcolor{Mycolor2}{\ding{70}} include RV data, but distinct sets of astrometric data, making the position angles effectively independent.
        The posterior represented by black contours include all available RV and astrometric data.
        We draw the readers attention in particular to the $\Omega$ vs. $i$ and $\Omega$ vs. $m$ sub-panels which illuminate how each data source progressively constrains the orientation of the orbit.}
    \label{fig:model-comparison}
\end{figure*}

\begin{figure*}
    \centering
    \hspace{-10pt}
    \includegraphics[width=0.95\linewidth]{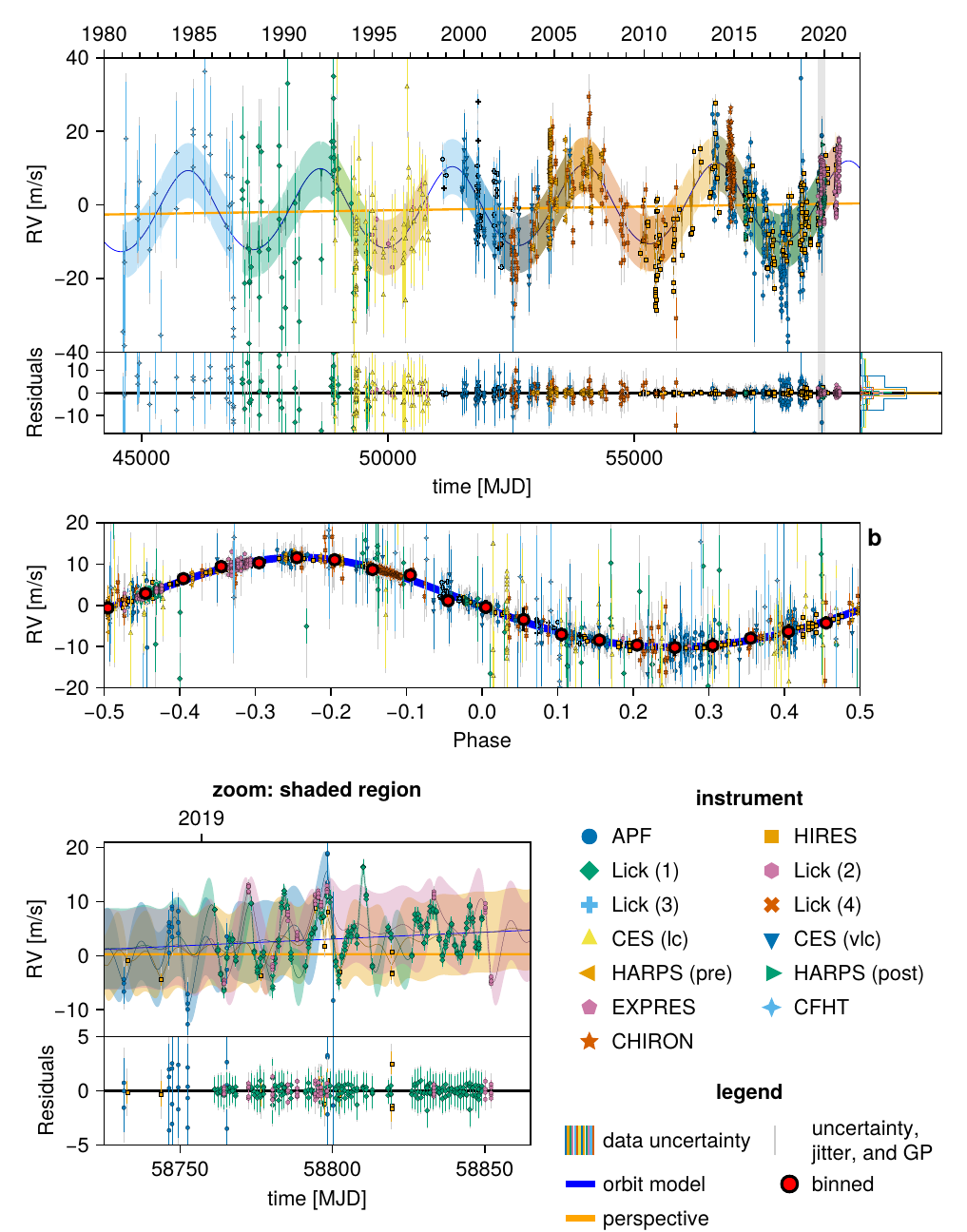}
    \caption{
        Maximum a-posteriori sample from the complete RV and astrometry model.
        Top: RV data from each instrument 
        The scatter in the RV data are well-modeled by a Gaussian process with a quasi-periodic kernel, just as reported in previous works.
    }
    \label{fig:rvpostplot}
\end{figure*}

\begin{figure}
    \begin{adjustwidth}{-1.8cm}{-0.3cm} 
    \includegraphics[width=\linewidth]{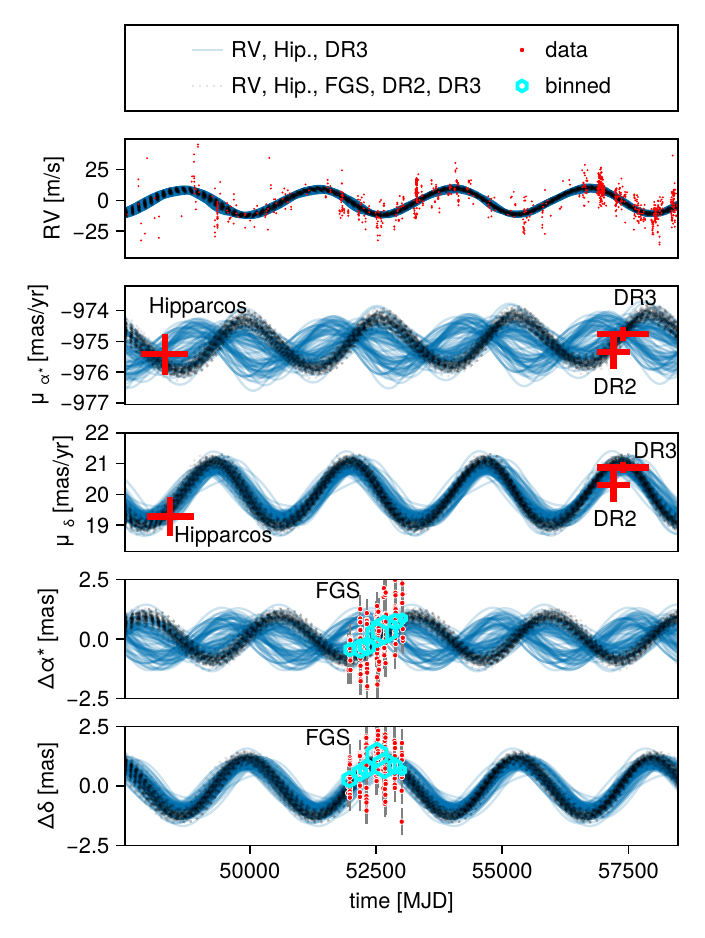}
    \end{adjustwidth}
    \caption{
        Orbits sampled from the RV, Hipparcos, DR3 model (blue) and the complete RV, Hippacos, FGS, DR2, DR3 model (black) compared against all data.
        This shows how the addition of the DR2 and the FGS data constrains the motion of the star in the R.A. direction.
        The horizontal bars in the proper motion panels show the time range over which each measurement is fit by their respective catalogs.
        The light blue markers show binned FGS data with the size of the marker indicating the number of data points in the bin.
        \label{fig:fgs-fit-comparison}}
\end{figure}

\begin{figure}
    \centering
    \begin{adjustwidth}{-1.0cm}{-0.45cm} 
    \includegraphics[width=\linewidth]{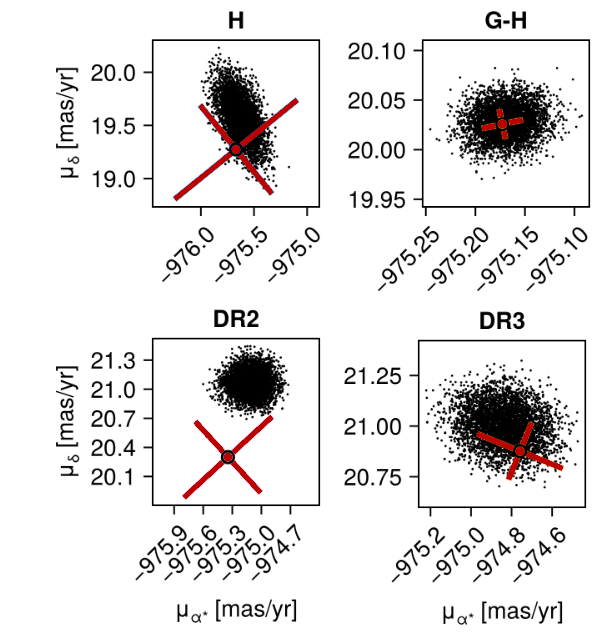}
    \end{adjustwidth}
    \caption{Covariances between R.A. and Dec. velocity at each proper motion epoch, for the RV, Hip., FGS, DR2, DR3 model. The "H" epoch is the net proper motion at the Hipparcos epoch calibrated to the DR3 velocity reference by the HGCA, "G-H" is the long-term proper motion derived from the HGCA, "DR2" is the Gaia DR2 reported proper motion after correcting for frame rotation, and "DR3" is the Gaia DR3 proper motion. Both the DR2 and DR3 values have uncertainties inflated according to the multipliers given in the HGCA, DR2 \citep{hgca_brandt_2018} and eDR3 versions \citep{hgca_brandt_2021}.}
    \label{fig:resids}
\end{figure}

\section{Results}

We now present the results of our models, starting with only the RV data and building up to the final solution.
The posteriors from all seven models are compared in Figure \ref{fig:orbit-comparison} and \ref{fig:model-comparison}, which includes a corner plot and predictions for 2025-01-01. We use consistent colors to identify each model throughout the figures in this section.
We report results using a mix of credible intervals (CI) and highest density intervals (HDI) as appropriate to summarize peaked or truncated distributions, with the $\pm$ notation used to indicate the median and 14\% / 84\% quantiles, and the [lower,upper] notation to indicate the the 75\% HDI interval. As with other works, we use 75\% to emphasize the arbitrariness of the specific probability mass fraction. 
Tables summarizing the posteriors of each model are presented in the Appendix. 
The final posterior corresponding to model 7 is available by request.

\subsection{RV Data}

Our model of the RV data alone results in a posterior that is unimodal, with a well-constrained semi-major axis of $3.54 \pm 0.04$ AU and a semi-amplitude of $10.6$ m/s, which is consistent with previous works.
The maximum \emph{a-posteriori} orbit sample from the RV-only posterior is displayed in Figure \ref{fig:rvpostplot}.
The orbital phase is well determined. The RV data and parallax alone show that the planet will be at opposition (its maximum separation from the star) in 2017, 2021, 2025, 2029, and so on. During these times, the planet will be localized to a ring of $1071\pm58$ mas separation.
The minimum mass, $m \sin(i)$, is $0.63 \pm 0.05 \,\mjup$.
The RV data supports an eccentricity that is consistent with 0: the 75\% highest density interval is $[0, 0.11]$.

The GP hyper-parameters adopt values similar to previous works, with a rotation period, $P_{rot}$, of $11.1 \pm0.5$ d, and decay timescale, $B$, of $57 \pm 5$ d.

\subsection{Constraints added by the Hipparcos IAD}

Adding the Hipparcos IAD data provides constraints on the inclination, position angle of ascending node, and position angle of the planet versus time.
This constrains the planet to be roughly on the South side of the star in 2025, North in 2029, and so on.
It provides essentially no information on the horizontal position of the planet.
The addition of the Hipparcos IAD does not significantly constrain the mass of the planet, other than that it is less than 1.3 \mjup\ (84 percentile upper limit), or the 75\% HDI of [0.56, 1.09] \mjup.

In the $i$ versus $\Omega$ subpanel of Figure \ref{fig:model-comparison}, we see that the Hipparcos IAD data does not provide any strong constraints on the inclination of the planet's orbit, but it does suggest that if the planet is co-planar with the outer disk, then it is more likely to orbiting counterclockwise.

\subsection{Constraints added by the HGCA}

Adding the DR3 Hipparcos-Gaia proper motion anomaly from the eDR3 HGCA \citep{hgca_brandt_2021} to the RV data provides significantly stronger constraints.
Note that this model includes only net velocities from the Gaia and Hipparcos cataloges, and does not include the within-catalogue motion from the Hipparcos IAD.
With RV and the HGCA data, we find a mass upper limit (84th percentile) of 1.00 \mjup.
With regard to the orientation of the orbit, the HGCA data constrains the inclination to a bimodal distribution of either $53\pm14\,^\circ$ or $130\pm13\,^\circ$, and a position angle of ascending node of $198\pm16\,^\circ$, both of which are in agreement with the orientation of the outer debris disk to $1.5\sigma$ \citep{epseri_alma_booth_2023}.

From the RV and HGCA data, we can constrain the position angle of the planet to be SSW of the star in 2025 in a region only about 600mas wide.

Figure \ref{fig:fgs-fit-comparison} shows the RV, proper motion, and position perturbation in R.A. and Dec. versus time for different orbit draws.
Looking at the proper motion panels, the orbit is near its most extreme values in the Dec. direction, but at mid-phase in R.A.---hence why the motion in Dec. is well-determined but ambiguous in R.A. with these data alone.

\subsection{Constraints added by the Hubble FGS data}

Adding the Hubble FGS data from \citet{fgs_eps_eri_benedict_2006} independently constrains the planet to be SSE or S of the star in 2025. 
These data prefer a slightly lower 84\% upper mass limit of 0.89 \mjup.

Looking at Figure \ref{fig:fgs-fit-comparison}, we see that the addition of the FGS data to the RV and HGCA removes an ambiguity in R.A. motion of the star, while agreeing well with the Dec. motion already determined by the HGCA. 

We see that the curvature in the FGS data is in very good agreement with the sinusoidal model, though recall that the absolute value and linear trend are fit flexibly as part of the model.

\subsection{Constraints added by the Gaia DR3-DR2 acceleration}
The addition of the DR2 data acts to significantly remove the ambiguity in the otherwise bimodal inclination marginal inclination posterior from the RV, Hipparcos, and DR3 model (employing the HGCA).
The result is similar to the addition of the FGS data, in that it acts to removes an ambiguity in the R.A. motion of the star. 
Notably, the constraint provided by either the DR3-DR2 acceleration \emph{or} the Hubble FGS astrometry selects the same part of the ambiguous posterior from the RV \& HGCA alone, increasing our confidence in the model.

\subsection{Complete Model}

With the complete model, the planet is very well-localized (Figure \ref{fig:model-comparison}, top right panel).
The model yields a  semi-major axis of $3.53\pm0.04$ AU, a mass of $1.00\pm0.10\,\mjup$, eccentricity of [0.00,0.10], and an orbit that is inclined $40^{+6}_{-5}\,^\circ$.
The direction of motion is counterclockwise in the plane of the sky.

We find that all proper motion data are well-fit by the model, as shown in Figure \ref{fig:resids}.

\subsection{Astrometry only orbit solution}
The data from Hipparcos, Hubble FGS, Gaia DR2, and DR3 is sufficient to, for the first time, fit an astrometry-only model, without any RV data. Using a uniform prior on the orbital period of 0.5--100 yr, we find an orbit posterior that is multi-modal with two prominent modes, one at 7.2 yr and one at 17 yr.
The 7.2 yr mode gives a constrained orbit that agrees very well with the RV only solution: consistent eccentricity, time of conjunction, $m\sin(i)$, and low to moderate eccentricity (HDI of [0, 0.42], peaked at 0). This mode is also consistent with the orientation of the outer disk.
The 17 yr mode on the other hand is inconsistent with the RV data. It has a  higher $m \sin(i)$ and is inconsistent with the orientation of the outer disk.
We further experimented with a full joint RV and astrometry model with period forced to be near the 17 yr mode identified from the astrometry. This resulted in a high eccentricity solution that was significantly disfavored versus the unconstrained RV and astrometry model according to the Bayes factor.
We therefore perform a broad cut of the posterior to select select the 7 yr model by adopting a uniform prior on the period of 5.36 to 9.36 yr (see Table \ref{tab:variables}).
The orbital period from the remaining astrometry only samples agrees within $1\sigma$ with the RV only model.
The times of conjunction and opposition are likewise moderately well constrained and in agreement.
The astrometry data provides a near-independent constraint on the mass of the planet to $1.1^{+0.5}_{-0.2}$, with the caveat of us selecting the mode that that agrees with the orientation of the disk and the RV period.

The posterior is bi-modal in position angle of ascending node, though both modes could agree with the orientation of the outer debris disk to $\approx1\sigma$ (the orientation of a  disk is ambiguous to $\pm90^\circ$ in inclination and $\pm180^\circ$ in position angle of ascending node with the current data).

\needspace{3em}
\section{Discussion}

\subsection{Orbital Period}

All of our orbit models that include RV data support an orbital period in the range of 7.3--7.45 years, with the complete model giving $7.33 \pm 0.08$ yr. 
This is very similar to the period of $7.31 \pm 0.06$ yr reported by \citet{epseri_jorge_2021}, though somewhat shorter than the $7.36\pm0.05$ yr  or  $7.6 \pm  0.01 $  yr periods reported by  \citet{epseri_epsindi_feng_2023} and by \citet{fgs_eps_eri_benedict_2006} respectively.
We believe this disagreement stems from differences in the noise model used to deal with stellar activity, which we discuss in Section \ref{sec:discussion-ecc}.

Our model of the absolute astrometry only, with no RV data, supports a period of $7.25\pm0.26$ years. This independent support for the orbital period should lay to rest concerns that the RV signature of \epserib\ may be an artifact of long-term stellar activity cycles.

Indeed, the perturbation orbit of the star has a projected semi-major axis of 2.7 mas  which is just over twice the 1.1 mas apparent radius of the star (using a 0.738  $M_\odot$ stellar radius from \citet{stellar_radii_rains_2020}).
Such a large displacement of the photocenter over a period of years cannot be explained by stellar activity.

\vspace{-0.2em}
\subsection{Orbital Phase}

The RV model we present determines the overall phase of the orbit and constrains the times of conjunction and opposition to approximately one month. Given the known parallax of the system, this means that the RV data alone can constrain the separation of the planet to ring whenever it is near opposition---in 2017, 2021, 2025, 2029, and so on.
This was noted as well in \citet{epseri_mawet_2019}.

The work by \citet{epseri_jorge_2021} added constraints from Hipparcos and Gaia DR2 and presented predictions for the planet's location at two imaging epochs. The astrometry data added constraints to the PA versus time, but the additional data broadened the constraints to the planet's separation compared to RV data alone. 
We hypothesize this may be a result of the two-step process in which the RV data were modeled first, a kernel density estimate (KDE) was fit to the posterior, and then the KDE was used as a prior in the subsequent absolute astrometry fit. The variables used in the astrometry fit do not include the time of conjunction directly, expressing it instead as a combination of other variables. It could be that while the KDE accurately represented the marginal posteriors of the period, eccentricity, argument of periastron, and epoch of periastron passage, the detailed correlation structure may not have been captured with sufficient detail to give a tightly constrained prior on time of conjunction derived from a combination of the above.

Joint models like the ones presented here and in previous works \citet{fgs_eps_eri_benedict_2006,HipparcosIAD_planets_Reffert_2011,fgs_eps_eri_benedict_2020,epseri_epsindi_feng_2023} sidestep this challenge. 
This could explain why our RV and Hipparcos model has considerably less uncertainty on the separation of the planet near the times of conjunction (e.g. in 2017-09-01, see Figure \ref{fig:predictions}).

\needspace{6em}
\subsection{Eccentricity}\label{sec:discussion-ecc}

The eccentricity of the planet is a key concern since it has a strong impact its separation from the star at opposition. 
Our models all support a low eccentricity that is consistent with zero.
This is consistent with previous works that included a GP model of stellar activity \citep{epseri_mawet_2019,epseri_jorge_2021,eps_eri_chiron_Giguere_2016} (eccentricity near zero) and is inconsistent with works that did not such as \citet{epseri_epsindi_feng_2023} and \citet{fgs_eps_eri_benedict_2020} (eccentricity of 0.26 and 0.16 respectively).

\needspace{6em}
\subsubsection{Including a GP results in a low eccentricity} 
To determine the source of this disagreement, we compared an RV model with a GP and an RV model without a GP.
As expected from earlier studies, the model without a GP arrives at an eccentricity of $\approx0.2$, while the model with a GP supports a low eccentricity consistent with zero.

\needspace{6em}
\subsubsection{The GP results are robust to the choice of data}

\begin{figure}
    \vspace{-0.2cm}
    \begin{adjustwidth}{-1.0cm}{-0.3cm} 
    \centering
    \includegraphics[width=\linewidth]{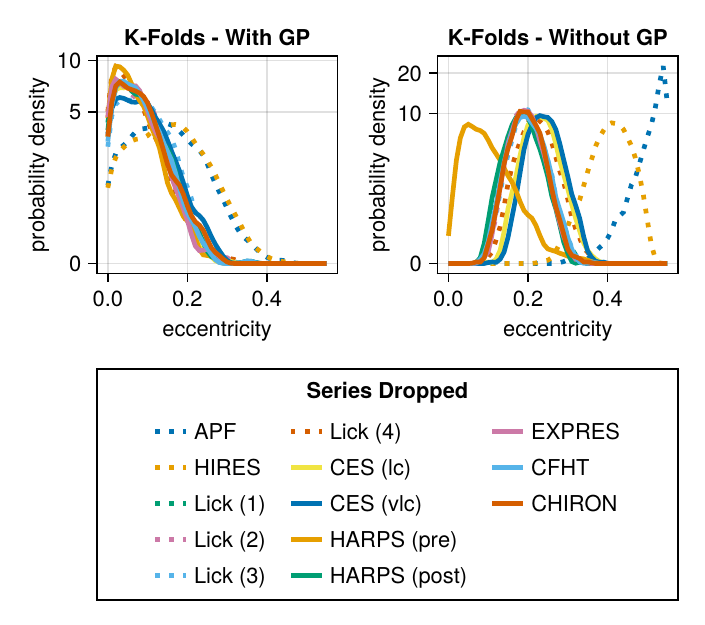}
    \end{adjustwidth}
    \caption{Marginal eccentricity posteriors from leave-one-instrument-out cross validation of RV datasets. The legend specifies which data set is \emph{ignored}. We find that the GP model (left) is more predictive of held out data than the model without a GP (right). Dropping the HARPS (pre) data from the no GP model also makes the eccentricity nearly consistent with zero ([0.02, 0.12])}
    \label{fig:ecc-cross-val}
\end{figure}
To understand why these two models arrive at different results, 
we performed two suites of leave-one-instrument-out cross validation.
For each model, we created 13 copies that each drop one of the instrument datasets (where e.g. Lick (1) and Lick (2) are treated as separate datasets). For this test only, we bounded the eccentricity to be at most 0.55.
We then sampled the posterior for all 26 models and compared their marginal eccentricity posteriors, shown in Figure \ref{fig:ecc-cross-val}.

We find that the GP model produces eccentricity posteriors that are  robust to the choice of dataset dropped. In all cases, we arrive at a probability distribution that is consistent with a near-circular orbit. The removal of either the APF or HIRES data broadens the eccentricity posterior but does not pull it significantly away from zero. This indicates that the APF and HIRES data provide the strongest upper bounds on the eccentricity, but does not indicate any tension from these datasets.

The no-GP model, on the other hand, is sensitive to the choice of dataset dropped. There are several datasets (HARPS (pre), Lick (4), and APF) that, when held out, lead to a drastic shift in the reported eccentricity.
Most notably, dropping the HARPS (pre) data results in an eccentricity posterior with a 75\% HDI of [0.035, 0.138] ($0.06 \pm 0.04$ CI),  compatible with zero.

While our main models used separate GPs for each instrument, we also tested a model with a single GP for all data. This setup, related to a strategy suggested in \citet{Blunt_2023}, shares temporal correlation information between instruments, instead of treating data from each instrument as uncorrelated.
We find essentially the same low-eccentricity results from this model, indicating that the GPs do not appear to be overfitting the data from each instrument.

\needspace{6em}
\subsubsection{High cadence data are the source of the disagreement}

\begin{figure}
    \begin{adjustwidth}{-1.25cm}{-0.5cm} 
    \centering
    \includegraphics[width=\linewidth]{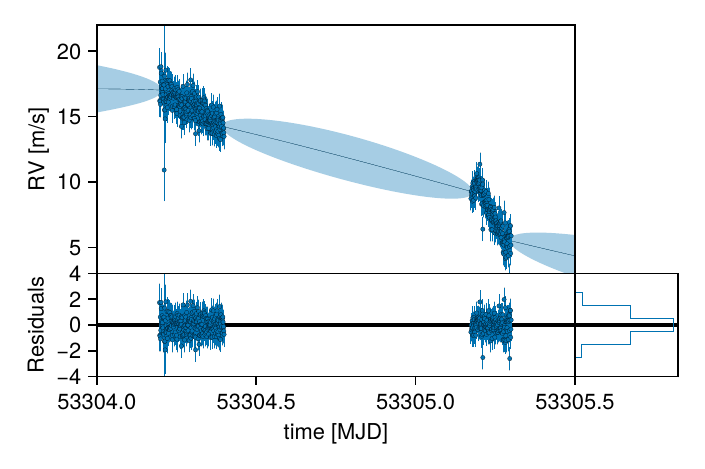}
    \end{adjustwidth}
    \caption{Zoom in of Figure \ref{fig:rvpostplot} showing the dense sampling of HARPS data over 1.5 days in October, 2004.
    The blue line and band show the GP model mean and standard deviation. The planet's contribution to the RV signal is essentially flat over this short time range.
    The inclusion of this data in particular causes an apparent eccentricity signal if not modeled with a GP. Dropping this data leads to a low eccentricity solution, even without a GP.}
    \label{fig:rvpostplot-crop}
\end{figure}

The HARPS (pre) series contains a burst of high cadence data, with 430 out of all 530 points captured over a 2-night range in October, 2004. This dense data from HARPS visibly traces out a correlated signal with a timescale much higher than the Keplerian signal (Figure \ref{fig:rvpostplot-crop}), which we interpret as a star-spot moving into or out of view following the 11.68 day rotation period of the star \citep{Barnes_2007}.
We find that dropping this data from the no-GP model leads to an eccentricity distribution that essentially matches that of our GP model.
The no-GP model is not equipped to recognize the high correlation between these points, and so the burst of high cadence data distorts the orbit model---producing both a non-zero eccentricity, and an overconfident posterior.

\needspace{6em}
\subsubsection{Possible impacts to previous RV models}

\citet{epseri_epsindi_feng_2023} used a moving average (MA) model to account for temporal correlation in the noise, which can be thought of as an simplified GP.

The MA model described in that work uses the past $q$ data points to predict the subset point, with $q=3$ for the HARPS (pre) data.
We hypothesize that the MA model may be affected in a similar way to the no-GP model when it encounters a burst of hundreds of data points over a short period of time. Earlier data points containing different realizations of the stellar noise may fall outside of this $q$-point memory, which could limit the models ability to recognize the temporal correlation and result in a higher eccentricity. It is possible that the MA model might arrive a low eccentricity when these points are dropped, just like our no-GP model does.

\subsubsection{Comparison with astrometry-only models}

It was noted by \citet{epseri_epsindi_feng_2023} that ``..astrometry-only analyses have also resulted in high-eccentricity solutions\dots  .'' While the cited works of \citet{fgs_eps_eri_benedict_2006} and \citet{fgs_eps_eri_benedict_2020} did arrive at moderate eccentricities, we should clarify that they did so using \emph{joint} models of RV and astrometry which likewise did not include a GP or other model of the stellar activity. 
In fact, no previous works have published an eccentricity value from an astrometry-only analysis to the best of our knowledge.
In contrast, our true astrometry-only model finds an eccentricity with a 75\% highest density interval of [0, 0.44], peaking at zero. The astrometry data therefore can be compatible with either the low or moderate eccentricity results.

\subsubsection{Conclusion: support for low eccentricity}

We find that the GP models do not appear to be over-fitting a real eccentricity signal, and in fact are \emph{more} predictive of held out data than the simpler model not including a GP.
Based on the improved predictive performance of the GP model, we conclude that the current data supports a low eccentricity orbit, consistent with zero to the precision of currently available data, in agreement with \citet{epseri_mawet_2019}, \citet{epseri_jorge_2021}, and \citet{epseri_express_Roettenbacher_2022}.

\subsection{Direction of Orbital Rotation}
We find that our models of individual astrometric data sources do not strongly constrain the direction of apparent orbital rotation in the plane of the sky to be clockwise or counterclockwise. 
For instance, the model of the Hipparcos and Gaia DR3 proper motions is evenly split between the two directions (inclination below/above $90^\circ$).
The model of the Hipparcos IAD does suggest that either the planet is moving counterclockwise and is near-coplanar with the disk, \emph{or}, that the planet is moving clockwise but misaligned with the disk.  
That said, the combination of all data sources, particularly the Hubble FGS and Gaia DR3-DR2 acceleration, supports counterclockwise motion fairly conclusively (see Figure \ref{fig:fgs-fit-comparison}). 
This result is inconsistent with that of \citet{epseri_epsindi_feng_2023} which is clockwise as indicated in their Figure 2.

\needspace{6em}
\subsection{Coplanarity with outer debris disk}

We now compare our complete orbit model to the outer debris disk model fit to ALMA data by \citet{epseri_alma_booth_2023}. Their posteriors indicate the disk has an inclination of $33.7 \pm 0.5 \, ^\circ$, and position angle of $-1.1 \pm 1.0 \, ^\circ$.
We find that the HGCA, Hubble FGS, and DR3-DR2 all support a planet that is co-planar with the outer debris disk.
Comparing the inclination of the complete model with that of the outer disk from \citet{epseri_alma_booth_2023}, we find a mutual inclination 75\% HDI of $[4,16]\,^\circ$.

It is worth noting that the uniform and sine priors we adopted for position angle of ascending node and inclination combine to produce an effective prior on the mutual inclination that goes to zero probability as the mutual inclination approaches zero. Given this, the mutual inclination we find is highly consistent with planet b being co-planar with the debis disk. Via the Savage–Dickey density ratio, we calculate a Bayes factor of approximately 130 for a model with a misalignment of $<2.5\,^\circ$, versus unconstrained mutual inclination. This is strong support in favor of alignment between the disk and planet.

We finally note that this conclusion is impacted by our inclusion of the DR2 proper motions. Although we have accounted for the bulk rotation between the DR2 and DR3 reference frames, there were additional changes to the processing of bright and high proper-motion stars between Gaia DR2 and (e)DR3 \citep{lindegren_2021_edr3_astrom}. These changes may result in an unknown shift between DR2 and DR3 that is unrelated to any underlying physical motion. The release of Gaia epoch astometry in a future data release promises to resolve this remaining question, as it will remove the need to compare data between Gaia releases.

\begin{figure}
    \centering
    \vspace{-0.3cm}
    \begin{adjustwidth}{-0.8cm}{-0.45cm} 
    \includegraphics[width=\linewidth]{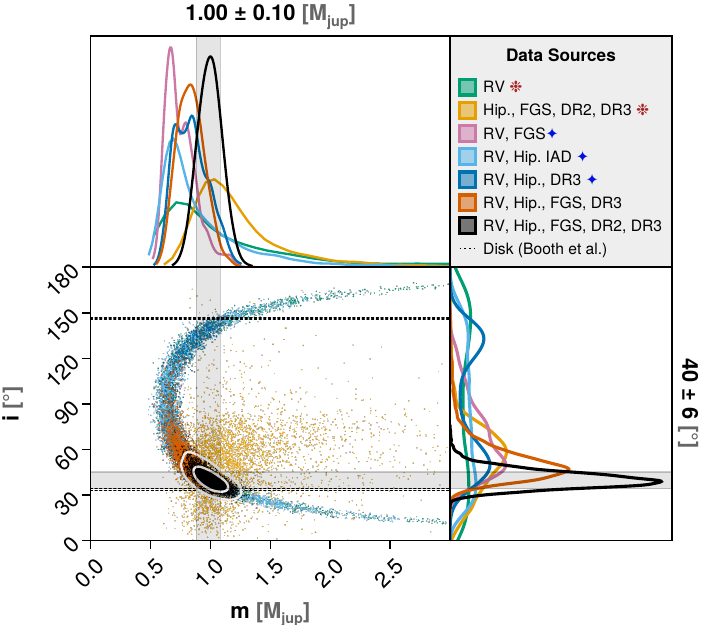}
    \end{adjustwidth}
    \caption{Marginal mass and inclination posteriors for planet b based on contributions from different datasets.
    We find that the model including all data, shown in black, finds a mass of $1.00\pm0.10 \,\mjup$ which is somewhat higher than previous estimates but in the range we would expect for a planet aligned with the outer debris disk.
    The white contours indicate $1\sigma$ and $3\sigma$, while the grey bands indicate the 68.5\% credible intervals.
    }
    \label{fig:mass-comparison}
\end{figure}

\subsection{Mass}

The RV data strongly constraints the $m \sin(i)$ of the planet, such that the mass we report is essentially determined only by the inclination of the orbit.
This situation is presented in Figure \ref{fig:mass-comparison}.
We find that the full model prefers an inclination of $41{+6}_{-5}\,^\circ$, which corresponds to a mass of $0.98^{+0.09}_{-0.10} \,\mjup$. If the planet is indeed perfectly coplanar with the outer disk to within $\pm1^\circ$,  this value would rise slightly to $1.04 \pm 0.06\;\mjup$. 

This mass is substantially higher than the value of $0.65^{+0.10}_{-0.09}\;\mjup$ reported by \citet{epseri_jorge_2021}, but this disagreement is only due to the different inclination ranges we find---adjusting the inclination of their posterior down to $40^\circ$ would give a mass of $1.00 \mjup$ as well. 

The mass we find is also higher than the value of $0.76^{+0.14}_{-0.11}\;\mjup$ reported by \citet{epseri_epsindi_feng_2023}, but in this case the disagreement comes partially from the different reported inclination ($130^\circ$) and partially from the reported non-zero eccentricity of 0.26.

Based on the revised mass, we calculated an expected contrast ratio of the planet versus star to be $2\times10^{-8}$ in L band, and $4\times10^{-6}$ in M band using Sonora Bobcat models \citep{sonora_bobcat} and an age of 439 Myr \citep{Barnes_2007}. That said, these are nominal values only as there is considerable uncertainty in the age of the system.

\vspace{-0.5em}
\subsection{Predictions}

\begin{figure*}
    \centering
    \includegraphics[width=\linewidth]{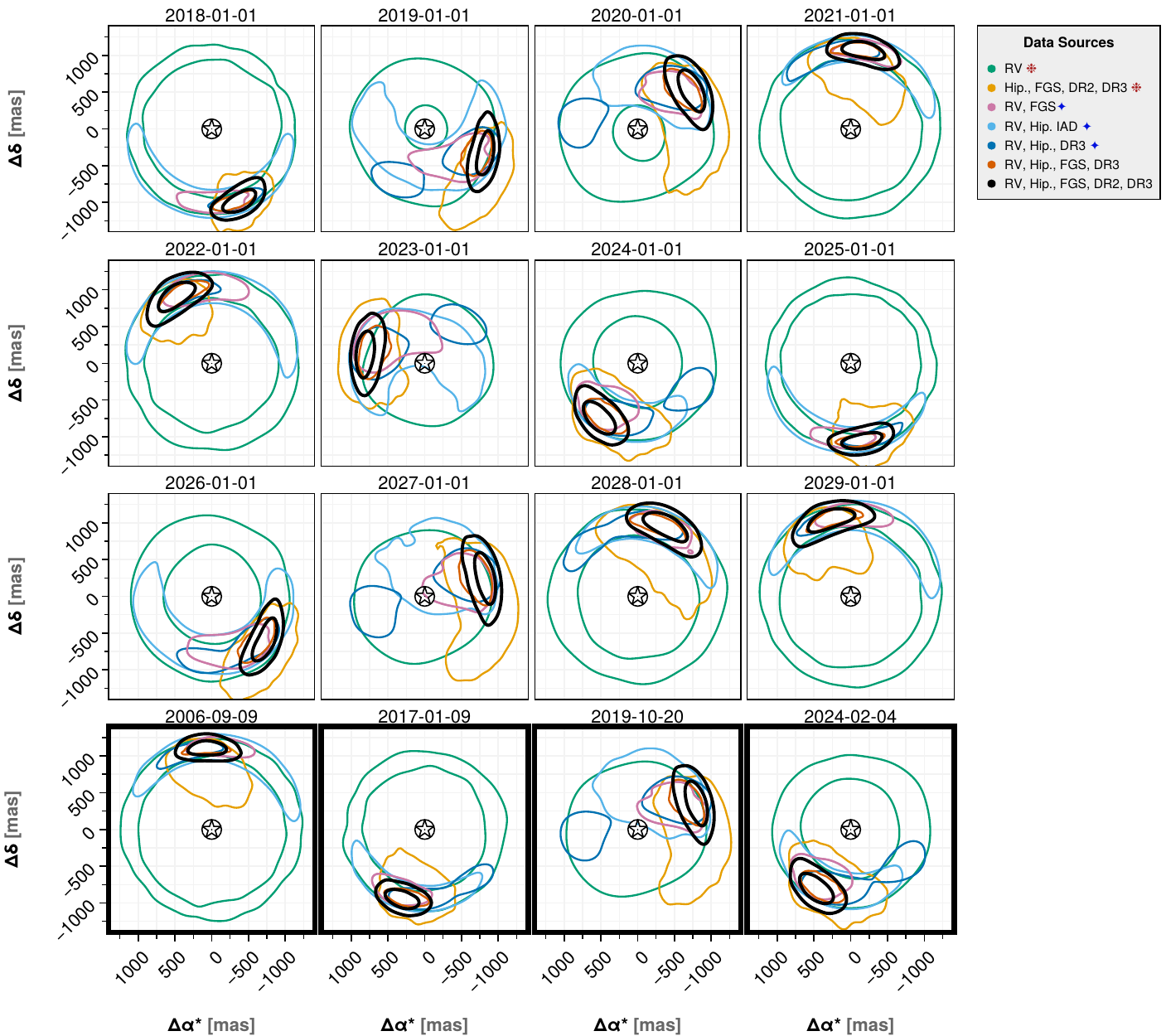}
    \caption{Predicted locations of planet b in one year increments. The contours represent $1\sigma$ Gaussian equivalent volumes. Both  $1\sigma$ and $3\sigma$ contours are plotted for the full model, indicated in black.
    The dark outlined panels in the bottom row correspond to notable imaging epochs.}
    \label{fig:predictions}
\end{figure*}

We present predictions in Figure \ref{fig:predictions} of the planet's apparent location for every year from 2018 through to 2029, in addition to four significant imaging epochs.
From the RV data alone, the planet's location is confined to a narrow ring of $1071\pm58$ mas at opposition, which occurs in 2024-12-07 + $N\times 3.66$ yr.
We the addition of any individual absolute astrometry dataset, the position angle of the planet at conjunction is confined to be South of the star in 2025, North in 2029, and so on. 
With our full model incorporating all absolute astrometry data sources, the position is confined to a $1\sigma$ region just $\approx 400 \times 250 $ mas wide near opposition.
With the revised inclination, an imaging detection of the planet away from opposition may also be feasible, as we find that it never appears closer to the star than $824\pm78$ mas.
This last constraint does rely on the complete model of all absolute astrometry, so conservative imaging programs may still wish to time their observations near opposition.

\subsection{Expected Constraints from Relative Astrometry}

With regards to the orbit and mass determination, the main benefit of an imaging detection will be an accurate measurement of the relative astrometry between the planet and star. 
While an imaging detection near opposition would be the most technically feasible, it would not contribute significant constraints to the inclination and therefore mass. It would nonetheless serve to confirm the orbit's position angle of ascending node.
To constrain inclination and mass, we recommend an imaging epoch somewhat away from opposition. The strongest constraints on the planet's mass would come from a relative astrometry measurement recorded close to a quarter period away from conjunction, i.e. when the planet is closest to the star. Such an endeavor would of course have to be balanced with the detectability at this closer separation.

\needspace{6em}
\section{Conclusions}

In this work we have presented an extensive reanalysis and cross-validation of all available radial velocity and absolute astrometry data. We determine the period of the planet's orbit to be $7.33 \pm 0.08$ yr ($3.53\pm 0.04$ AU). We  independently confirm the period using the first complete orbit model of the system from only absolute astrometry data. In doing so, we rule out the possibility that the 7.3 year periodic signal could be a result of stellar activity cycles instead of a planet.

We find a much lower overall inclination than previous works ($i=40^{+6}_{-5}\,^\circ$, $\Omega=186\pm9\,^\circ$) that is consistent with coplanarity with the outer debris disk ($33.7 \pm 0.5 \, ^\circ$, as derived from ALMA observations by \citep{epseri_alma_booth_2023}, with the 75\% HDI of the mutual inclination between $[2.5, 14]\,^\circ$.
The higher inclination we find supports a dynamical mass of $1.00 \pm 0.10 \,\mjup$.
We also determine that the planet's orbit has a low eccentricity of $[0,0.10]$ consistent with a near-circular orbit.

We discussed disagreements in the literature regarding the planet's time of conjunction and opposition; inclination; and eccentricity, and presented explanations for why our updated results differ.

Finally, we presented detailed predictions for the planet's apparent location versus time, broken down by different subsets of data. 

Our significantly revised mass and position predictions will aid future efforts to detect the planet through imaging, and point to \epserib\ being a remarkably close analog to our own planet Jupiter.

\begin{small}

\vspace{6em}
We thank T. Brandt for guidance in modeling the HGCA. We thank William Roberson for discussions regarding second order effects for high proper motion and barycentric radial velocity stars. We also thank Eric Agol for discussions regarding Celerite.jl and Gaussian processes. We thank Jessica Speedie for discussions regarding ALMA data. We thank Mark Booth for sharing posteriors from \citet{epseri_alma_booth_2023} in machine-readable format. 
We thank the anonymous referee for their insightful comments which have improved this work.

This work has made use of data from the European Space Agency (ESA) mission
{\it Gaia} (\url{https://www.cosmos.esa.int/gaia}), processed by the {\it Gaia}
Data Processing and Analysis Consortium (DPAC,
\url{https://www.cosmos.esa.int/web/gaia/dpac/consortium}). Funding for the DPAC
has been provided by national institutions, in particular the institutions
participating in the {\it Gaia} Multilateral Agreement.

\end{small}

\vspace{5mm}
\facilities{APF, CFHT, CTIO:1.5m (CHIRON), ESO:3.6m (CES), ESO:3.6m (HARPS), Gaia, HIPPARCOS, Keck:I (HIRES), LDT (EXPRES), VLT:Kueyen (UVES), HST (FGS).}


\software{Octofitter \citep{thompson_octofitter_2023}, Makie \citep{DanischKrumbiegel2021}, Julia \citep{bezanson_julia_2012}.}

\bibliography{ms}{}

\begin{thebibliography}{}
\expandafter\ifx\csname natexlab\endcsname\relax\def\natexlab#1{#1}\fi
\providecommand{\url}[1]{\href{#1}{#1}}
\providecommand{\dodoi}[1]{doi:~\href{http://doi.org/#1}{\nolinkurl{#1}}}
\providecommand{\doeprint}[1]{\href{http://ascl.net/#1}{\nolinkurl{http://ascl.net/#1}}}
\providecommand{\doarXiv}[1]{\href{https://arxiv.org/abs/#1}{\nolinkurl{https://arxiv.org/abs/#1}}}

\bibitem[{Barnes(2007)}]{Barnes_2007}
Barnes, S.~A. 2007, The Astrophysical Journal, 669, 1167, \dodoi{10.1086/519295}

\bibitem[{{Benedict} {et~al.}(2006){Benedict}, {McArthur}, {Gatewood}, {Nelan}, {Cochran}, {Hatzes}, {Endl}, {Wittenmyer}, {Baliunas}, {Walker}, {Yang}, {K{\"u}rster}, {Els}, \& {Paulson}}]{fgs_eps_eri_benedict_2006}
{Benedict}, G.~F., {McArthur}, B.~E., {Gatewood}, G., {et~al.} 2006, \aj, 132, 2206, \dodoi{10.1086/508323}

\bibitem[{{Bezanson} {et~al.}(2012){Bezanson}, {Karpinski}, {Shah}, \& {Edelman}}]{bezanson_julia_2012}
{Bezanson}, J., {Karpinski}, S., {Shah}, V.~B., \& {Edelman}, A. 2012, arXiv e-prints, arXiv:1209.5145, \dodoi{10.48550/arXiv.1209.5145}

\bibitem[{Blunt {et~al.}(2023)Blunt, Carvalho, David, Beichman, Zink, Gaidos, Behmard, Bouma, Cody, Dai, Foreman-Mackey, Grunblatt, Howard, Kosiarek, Knutson, Rubenzahl, Beard, Chontos, Giacalone, Hirano, Johnson, Lubin, Akana~Murphy, Petigura, Van~Zandt, \& Weiss}]{Blunt_2023}
Blunt, S., Carvalho, A., David, T.~J., {et~al.} 2023, The Astronomical Journal, 166, 62, \dodoi{10.3847/1538-3881/acde78}

\bibitem[{{Booth} {et~al.}(2023){Booth}, {Pearce}, {Krivov}, {Wyatt}, {Dent}, {Hales}, {Lestrade}, {Cruz-S{\'a}enz de Miera}, {Faramaz}, {L{\"o}hne}, \& {Chavez-Dagostino}}]{epseri_alma_booth_2023}
{Booth}, M., {Pearce}, T.~D., {Krivov}, A.~V., {et~al.} 2023, \mnras, 521, 6180, \dodoi{10.1093/mnras/stad938}

\bibitem[{{Brandt}(2018)}]{hgca_brandt_2018}
{Brandt}, T.~D. 2018, \apjs, 239, 31, \dodoi{10.3847/1538-4365/aaec06}

\bibitem[{{Brandt}(2021)}]{hgca_brandt_2021}
---. 2021, \apjs, 254, 42, \dodoi{10.3847/1538-4365/abf93c}

\bibitem[{{Campbell} {et~al.}(1988){Campbell}, {Walker}, \& {Yang}}]{rvs_cfht_campbell_1988}
{Campbell}, B., {Walker}, G.~A.~H., \& {Yang}, S. 1988, \apj, 331, 902, \dodoi{10.1086/166608}

\bibitem[{Danisch \& Krumbiegel(2021)}]{DanischKrumbiegel2021}
Danisch, S., \& Krumbiegel, J. 2021, Journal of Open Source Software, 6, 3349, \dodoi{10.21105/joss.03349}

\bibitem[{{Feng} {et~al.}(2023){Feng}, {Butler}, {Vogt}, {Holden}, \& {Rui}}]{epseri_epsindi_feng_2023}
{Feng}, F., {Butler}, R.~P., {Vogt}, S.~S., {Holden}, B., \& {Rui}, Y. 2023, \mnras, 525, 607, \dodoi{10.1093/mnras/stad2297}

\bibitem[{{Fischer} {et~al.}(2014){Fischer}, {Marcy}, \& {Spronck}}]{rvs_lick_fisher_2014}
{Fischer}, D.~A., {Marcy}, G.~W., \& {Spronck}, J. F.~P. 2014, \apjs, 210, 5, \dodoi{10.1088/0067-0049/210/1/5}

\bibitem[{{Foreman-Mackey} {et~al.}(2017){Foreman-Mackey}, {Agol}, {Ambikasaran}, \& {Angus}}]{celerite_dfm}
{Foreman-Mackey}, D., {Agol}, E., {Ambikasaran}, S., \& {Angus}, R. 2017, {celerite: Scalable 1D Gaussian Processes in C++, Python, and Julia}, Astrophysics Source Code Library, record ascl:1709.008

\bibitem[{Fritz~Benedict(2022)}]{fgs_eps_eri_benedict_2020}
Fritz~Benedict, G. 2022, Research Notes of the AAS, 6, 45, \dodoi{10.3847/2515-5172/ac5b6b}

\bibitem[{{Gaia Collaboration} {et~al.}(2016){Gaia Collaboration}, {Prusti}, {de Bruijne}, {Brown}, {Vallenari}, {Babusiaux}, {Bailer-Jones}, {Bastian}, {Biermann}, {Evans}, {Eyer}, {Jansen}, {Jordi}, {Klioner}, {Lammers}, {Lindegren}, {Luri}, {Mignard}, {Milligan}, {Panem}, {Poinsignon}, {Pourbaix}, {Randich}, {Sarri}, {Sartoretti}, {Siddiqui}, {Soubiran}, {Valette}, {van Leeuwen}, {Walton}, {Aerts}, {Arenou}, {Cropper}, {Drimmel}, {H{\o}g}, {Katz}, {Lattanzi}, {O'Mullane}, {Grebel}, {Holland}, {Huc}, {Passot}, {Bramante}, {Cacciari}, {Casta{\~n}eda}, {Chaoul}, {Cheek}, {De Angeli}, {Fabricius}, {Guerra}, {Hern{\'a}ndez}, {Jean-Antoine-Piccolo}, {Masana}, {Messineo}, {Mowlavi}, {Nienartowicz}, {Ord{\'o}{\~n}ez-Blanco}, {Panuzzo}, {Portell}, {Richards}, {Riello}, {Seabroke}, {Tanga}, {Th{\'e}venin}, {Torra}, {Els}, {Gracia-Abril}, {Comoretto}, {Garcia-Reinaldos}, {Lock}, {Mercier}, {Altmann}, {Andrae}, {Astraatmadja}, {Bellas-Velidis}, {Benson}, {Berthier}, {Blomme}, {Busso}, {Carry}, {Cellino}, {Clementini},
  {Cowell}, {Creevey}, {Cuypers}, {Davidson}, {De Ridder}, {de Torres}, {Delchambre}, {Dell'Oro}, {Ducourant}, {Fr{\'e}mat}, {Garc{\'\i}a-Torres}, {Gosset}, {Halbwachs}, {Hambly}, {Harrison}, {Hauser}, {Hestroffer}, {Hodgkin}, {Huckle}, {Hutton}, {Jasniewicz}, {Jordan}, {Kontizas}, {Korn}, {Lanzafame}, {Manteiga}, {Moitinho}, {Muinonen}, {Osinde}, {Pancino}, {Pauwels}, {Petit}, {Recio-Blanco}, {Robin}, {Sarro}, {Siopis}, {Smith}, {Smith}, {Sozzetti}, {Thuillot}, {van Reeven}, {Viala}, {Abbas}, {Abreu Aramburu}, {Accart}, {Aguado}, {Allan}, {Allasia}, {Altavilla}, {{\'A}lvarez}, {Alves}, {Anderson}, {Andrei}, {Anglada Varela}, {Antiche}, {Antoja}, {Ant{\'o}n}, {Arcay}, {Atzei}, {Ayache}, {Bach}, {Baker}, {Balaguer-N{\'u}{\~n}ez}, {Barache}, {Barata}, {Barbier}, {Barblan}, {Baroni}, {Barrado y Navascu{\'e}s}, {Barros}, {Barstow}, {Becciani}, {Bellazzini}, {Bellei}, {Bello Garc{\'\i}a}, {Belokurov}, {Bendjoya}, {Berihuete}, {Bianchi}, {Bienaym{\'e}}, {Billebaud}, {Blagorodnova}, {Blanco-Cuaresma}, {Boch},
  {Bombrun}, {Borrachero}, {Bouquillon}, {Bourda}, {Bouy}, {Bragaglia}, {Breddels}, {Brouillet}, {Br{\"u}semeister}, {Bucciarelli}, {Budnik}, {Burgess}, {Burgon}, {Burlacu}, {Busonero}, {Buzzi}, {Caffau}, {Cambras}, {Campbell}, {Cancelliere}, {Cantat-Gaudin}, {Carlucci}, {Carrasco}, {Castellani}, {Charlot}, {Charnas}, {Charvet}, {Chassat}, {Chiavassa}, {Clotet}, {Cocozza}, {Collins}, {Collins}, \& {Costigan}}]{gaia}
{Gaia Collaboration}, {Prusti}, T., {de Bruijne}, J.~H.~J., {et~al.} 2016, \aap, 595, A1, \dodoi{10.1051/0004-6361/201629272}

\bibitem[{{Gaia Collaboration} {et~al.}(2018){Gaia Collaboration}, {Brown}, {Vallenari}, {Prusti}, {de Bruijne}, {Babusiaux}, {Bailer-Jones}, {Biermann}, {Evans}, {Eyer}, {Jansen}, {Jordi}, {Klioner}, {Lammers}, {Lindegren}, {Luri}, {Mignard}, {Panem}, {Pourbaix}, {Randich}, {Sartoretti}, {Siddiqui}, {Soubiran}, {van Leeuwen}, {Walton}, {Arenou}, {Bastian}, {Cropper}, {Drimmel}, {Katz}, {Lattanzi}, {Bakker}, {Cacciari}, {Casta{\~n}eda}, {Chaoul}, {Cheek}, {De Angeli}, {Fabricius}, {Guerra}, {Holl}, {Masana}, {Messineo}, {Mowlavi}, {Nienartowicz}, {Panuzzo}, {Portell}, {Riello}, {Seabroke}, {Tanga}, {Th{\'e}venin}, {Gracia-Abril}, {Comoretto}, {Garcia-Reinaldos}, {Teyssier}, {Altmann}, {Andrae}, {Audard}, {Bellas-Velidis}, {Benson}, {Berthier}, {Blomme}, {Burgess}, {Busso}, {Carry}, {Cellino}, {Clementini}, {Clotet}, {Creevey}, {Davidson}, {De Ridder}, {Delchambre}, {Dell'Oro}, {Ducourant}, {Fern{\'a}ndez-Hern{\'a}ndez}, {Fouesneau}, {Fr{\'e}mat}, {Galluccio}, {Garc{\'\i}a-Torres},
  {Gonz{\'a}lez-N{\'u}{\~n}ez}, {Gonz{\'a}lez-Vidal}, {Gosset}, {Guy}, {Halbwachs}, {Hambly}, {Harrison}, {Hern{\'a}ndez}, {Hestroffer}, {Hodgkin}, {Hutton}, {Jasniewicz}, {Jean-Antoine-Piccolo}, {Jordan}, {Korn}, {Krone-Martins}, {Lanzafame}, {Lebzelter}, {L{\"o}ffler}, {Manteiga}, {Marrese}, {Mart{\'\i}n-Fleitas}, {Moitinho}, {Mora}, {Muinonen}, {Osinde}, {Pancino}, {Pauwels}, {Petit}, {Recio-Blanco}, {Richards}, {Rimoldini}, {Robin}, {Sarro}, {Siopis}, {Smith}, {Sozzetti}, {S{\"u}veges}, {Torra}, {van Reeven}, {Abbas}, {Abreu Aramburu}, {Accart}, {Aerts}, {Altavilla}, {{\'A}lvarez}, {Alvarez}, {Alves}, {Anderson}, {Andrei}, {Anglada Varela}, {Antiche}, {Antoja}, {Arcay}, {Astraatmadja}, {Bach}, {Baker}, {Balaguer-N{\'u}{\~n}ez}, {Balm}, {Barache}, {Barata}, {Barbato}, {Barblan}, {Barklem}, {Barrado}, {Barros}, {Barstow}, {Bartholom{\'e} Mu{\~n}oz}, {Bassilana}, {Becciani}, {Bellazzini}, {Berihuete}, {Bertone}, {Bianchi}, {Bienaym{\'e}}, {Blanco-Cuaresma}, {Boch}, {Boeche}, {Bombrun}, {Borrachero},
  {Bossini}, {Bouquillon}, {Bourda}, {Bragaglia}, {Bramante}, {Breddels}, {Bressan}, {Brouillet}, {Br{\"u}semeister}, {Brugaletta}, {Bucciarelli}, {Burlacu}, {Busonero}, {Butkevich}, {Buzzi}, {Caffau}, {Cancelliere}, {Cannizzaro}, {Cantat-Gaudin}, {Carballo}, {Carlucci}, {Carrasco}, {Casamiquela}, {Castellani}, {Castro-Ginard}, {Charlot}, {Chemin}, {Chiavassa}, {Cocozza}, {Costigan}, {Cowell}, {Crifo}, {Crosta}, {Crowley}, {Cuypers}, {Dafonte}, {Damerdji}, {Dapergolas}, {David}, {David}, {de Laverny}, \& {De Luise}}]{gaia_dr2}
{Gaia Collaboration}, {Brown}, A.~G.~A., {Vallenari}, A., {et~al.} 2018, \aap, 616, A1, \dodoi{10.1051/0004-6361/201833051}

\bibitem[{{Gaia Collaboration} {et~al.}(2023){Gaia Collaboration}, {Vallenari}, {Brown}, {Prusti}, {de Bruijne}, {Arenou}, {Babusiaux}, {Biermann}, {Creevey}, {Ducourant}, {Evans}, {Eyer}, {Guerra}, {Hutton}, {Jordi}, {Klioner}, {Lammers}, {Lindegren}, {Luri}, {Mignard}, {Panem}, {Pourbaix}, {Randich}, {Sartoretti}, {Soubiran}, {Tanga}, {Walton}, {Bailer-Jones}, {Bastian}, {Drimmel}, {Jansen}, {Katz}, {Lattanzi}, {van Leeuwen}, {Bakker}, {Cacciari}, {Casta{\~n}eda}, {De Angeli}, {Fabricius}, {Fouesneau}, {Fr{\'e}mat}, {Galluccio}, {Guerrier}, {Heiter}, {Masana}, {Messineo}, {Mowlavi}, {Nicolas}, {Nienartowicz}, {Pailler}, {Panuzzo}, {Riclet}, {Roux}, {Seabroke}, {Sordo}, {Th{\'e}venin}, {Gracia-Abril}, {Portell}, {Teyssier}, {Altmann}, {Andrae}, {Audard}, {Bellas-Velidis}, {Benson}, {Berthier}, {Blomme}, {Burgess}, {Busonero}, {Busso}, {C{\'a}novas}, {Carry}, {Cellino}, {Cheek}, {Clementini}, {Damerdji}, {Davidson}, {de Teodoro}, {Nu{\~n}ez Campos}, {Delchambre}, {Dell'Oro}, {Esquej},
  {Fern{\'a}ndez-Hern{\'a}ndez}, {Fraile}, {Garabato}, {Garc{\'\i}a-Lario}, {Gosset}, {Haigron}, {Halbwachs}, {Hambly}, {Harrison}, {Hern{\'a}ndez}, {Hestroffer}, {Hodgkin}, {Holl}, {Jan{\ss}en}, {Jevardat de Fombelle}, {Jordan}, {Krone-Martins}, {Lanzafame}, {L{\"o}ffler}, {Marchal}, {Marrese}, {Moitinho}, {Muinonen}, {Osborne}, {Pancino}, {Pauwels}, {Recio-Blanco}, {Reyl{\'e}}, {Riello}, {Rimoldini}, {Roegiers}, {Rybizki}, {Sarro}, {Siopis}, {Smith}, {Sozzetti}, {Utrilla}, {van Leeuwen}, {Abbas}, {{\'A}brah{\'a}m}, {Abreu Aramburu}, {Aerts}, {Aguado}, {Ajaj}, {Aldea-Montero}, {Altavilla}, {{\'A}lvarez}, {Alves}, {Anders}, {Anderson}, {Anglada Varela}, {Antoja}, {Baines}, {Baker}, {Balaguer-N{\'u}{\~n}ez}, {Balbinot}, {Balog}, {Barache}, {Barbato}, {Barros}, {Barstow}, {Bartolom{\'e}}, {Bassilana}, {Bauchet}, {Becciani}, {Bellazzini}, {Berihuete}, {Bernet}, {Bertone}, {Bianchi}, {Binnenfeld}, {Blanco-Cuaresma}, {Blazere}, {Boch}, {Bombrun}, {Bossini}, {Bouquillon}, {Bragaglia}, {Bramante}, {Breedt},
  {Bressan}, {Brouillet}, {Brugaletta}, {Bucciarelli}, {Burlacu}, {Butkevich}, {Buzzi}, {Caffau}, {Cancelliere}, {Cantat-Gaudin}, {Carballo}, {Carlucci}, {Carnerero}, {Carrasco}, {Casamiquela}, {Castellani}, {Castro-Ginard}, {Chaoul}, {Charlot}, {Chemin}, {Chiaramida}, {Chiavassa}, {Chornay}, {Comoretto}, {Contursi}, {Cooper}, {Cornez}, {Cowell}, {Crifo}, {Cropper}, {Crosta}, {Crowley}, {Dafonte}, {Dapergolas}, {David}, {David}, {de Laverny}, {De Luise}, \& {De March}}]{gaia_dr3}
{Gaia Collaboration}, {Vallenari}, A., {Brown}, A.~G.~A., {et~al.} 2023, \aap, 674, A1, \dodoi{10.1051/0004-6361/202243940}

\bibitem[{{Gatewood}(2000)}]{epseri_proceeding_gatewoord_2000}
{Gatewood}, G. 2000, in AAS/Division for Planetary Sciences Meeting Abstracts, Vol.~32, AAS/Division for Planetary Sciences Meeting Abstracts \#32, 32.01

\bibitem[{{Gatewood}(1987)}]{map_astrometry_gatewood_1987}
{Gatewood}, G.~D. 1987, \aj, 94, 213, \dodoi{10.1086/114466}

\bibitem[{Gelman \& Rubin(1992)}]{rhat_gelman}
Gelman, A., \& Rubin, D.~B. 1992, Statistical Science, 7, 457 , \dodoi{10.1214/ss/1177011136}

\bibitem[{Giguere {et~al.}(2016)Giguere, Fischer, Zhang, Matthews, Cameron, \& Henry}]{eps_eri_chiron_Giguere_2016}
Giguere, M.~J., Fischer, D.~A., Zhang, C. X.~Y., {et~al.} 2016, The Astrophysical Journal, 824, 150, \dodoi{10.3847/0004-637X/824/2/150}

\bibitem[{Hatzes {et~al.}(2000)Hatzes, Cochran, McArthur, Baliunas, Walker, Campbell, Irwin, Yang, Kürster, Endl, Els, Butler, \& Marcy}]{epseri_lick_Hatzes_2000}
Hatzes, A.~P., Cochran, W.~D., McArthur, B., {et~al.} 2000, The Astrophysical Journal, 544, L145, \dodoi{10.1086/317319}

\bibitem[{{Heinze} {et~al.}(2008){Heinze}, {Hinz}, {Kenworthy}, {Miller}, \& {Sivanandam}}]{epseri_imaging_mmt_heinze_2008}
{Heinze}, A.~N., {Hinz}, P.~M., {Kenworthy}, M., {Miller}, D., \& {Sivanandam}, S. 2008, \apj, 688, 583, \dodoi{10.1086/592100}

\bibitem[{{Janson} {et~al.}(2008){Janson}, {Reffert}, {Brandner}, {Henning}, {Lenzen}, \& {Hippler}}]{epseri_janson_2008}
{Janson}, M., {Reffert}, S., {Brandner}, W., {et~al.} 2008, \aap, 488, 771, \dodoi{10.1051/0004-6361:200809984}

\bibitem[{{Janson, Markus} {et~al.}(2015){Janson, Markus}, {Quanz, Sascha P.}, {Carson, Joseph C.}, {Thalmann, Christian}, {Lafrenière, David}, \& {Amara, Adam}}]{epseri_janson_2015}
{Janson, Markus}, {Quanz, Sascha P.}, {Carson, Joseph C.}, {et~al.} 2015, A\&A, 574, A120, \dodoi{10.1051/0004-6361/201424944}

\bibitem[{{Lindegren} {et~al.}(2021){Lindegren}, {Klioner}, {Hern{\'a}ndez}, {Bombrun}, {Ramos-Lerate}, {Steidelm{\"u}ller}, {Bastian}, {Biermann}, {de Torres}, {Gerlach}, {Geyer}, {Hilger}, {Hobbs}, {Lammers}, {McMillan}, {Stephenson}, {Casta{\~n}eda}, {Davidson}, {Fabricius}, {Gracia-Abril}, {Portell}, {Rowell}, {Teyssier}, {Torra}, {Bartolom{\'e}}, {Clotet}, {Garralda}, {Gonz{\'a}lez-Vidal}, {Torra}, {Abbas}, {Altmann}, {Anglada Varela}, {Balaguer-N{\'u}{\~n}ez}, {Balog}, {Barache}, {Becciani}, {Bernet}, {Bertone}, {Bianchi}, {Bouquillon}, {Brown}, {Bucciarelli}, {Busonero}, {Butkevich}, {Buzzi}, {Cancelliere}, {Carlucci}, {Charlot}, {Cioni}, {Crosta}, {Crowley}, {del Peloso}, {del Pozo}, {Drimmel}, {Esquej}, {Fienga}, {Fraile}, {Gai}, {Garcia-Reinaldos}, {Guerra}, {Hambly}, {Hauser}, {Jan{\ss}en}, {Jordan}, {Kostrzewa-Rutkowska}, {Lattanzi}, {Liao}, {Licata}, {Lister}, {L{\"o}ffler}, {Marchant}, {Masip}, {Mignard}, {Mints}, {Molina}, {Mora}, {Morbidelli}, {Murphy}, {Pagani}, {Panuzzo}, {Pe{\~n}alosa
  Esteller}, {Poggio}, {Re Fiorentin}, {Riva}, {Sagrist{\`a} Sell{\'e}s}, {Sanchez Gimenez}, {Sarasso}, {Sciacca}, {Siddiqui}, {Smart}, {Souami}, {Spagna}, {Steele}, {Taris}, {Utrilla}, {van Reeven}, \& {Vecchiato}}]{lindegren_2021_edr3_astrom}
{Lindegren}, L., {Klioner}, S.~A., {Hern{\'a}ndez}, J., {et~al.} 2021, \aap, 649, A2, \dodoi{10.1051/0004-6361/202039709}

\bibitem[{{Lindegren, Lennart}(2020)}]{dr2_reference_frame_Lindegren_2020}
{Lindegren, Lennart}. 2020, A\&A, 633, A1, \dodoi{10.1051/0004-6361/201936161}

\bibitem[{{Llop-Sayson} {et~al.}(2021){Llop-Sayson}, {Wang}, {Ruffio}, {Mawet}, {Blunt}, {Absil}, {Bond}, {Brinkman}, {Bowler}, {Bottom}, {Chontos}, {Dalba}, {Fulton}, {Giacalone}, {Hill}, {Hirsch}, {Howard}, {Isaacson}, {Karlsson}, {Lubin}, {Madurowicz}, {Matthews}, {Morris}, {Perrin}, {Ren}, {Rice}, {Rosenthal}, {Ruane}, {Rubenzahl}, {Sun}, {Wallack}, {Xuan}, \& {Ygouf}}]{epseri_jorge_2021}
{Llop-Sayson}, J., {Wang}, J.~J., {Ruffio}, J.-B., {et~al.} 2021, \aj, 162, 181, \dodoi{10.3847/1538-3881/ac134a}

\bibitem[{{Macintosh} {et~al.}(2003){Macintosh}, {Becklin}, {Kaisler}, {Konopacky}, \& {Zuckerman}}]{epseri_kband_macintosh_2003}
{Macintosh}, B.~A., {Becklin}, E.~E., {Kaisler}, D., {Konopacky}, Q., \& {Zuckerman}, B. 2003, \apj, 594, 538, \dodoi{10.1086/376827}

\bibitem[{{Mamajek} \& {Hillenbrand}(2008)}]{mamajek_2008}
{Mamajek}, E.~E., \& {Hillenbrand}, L.~A. 2008, \apj, 687, 1264, \dodoi{10.1086/591785}

\bibitem[{Marley {et~al.}(2021)Marley, Saumon, Morley, Fortney, Visscher, Freedman, \& Lupu}]{sonora_bobcat}
Marley, M., Saumon, D., Morley, C., {et~al.} 2021, Sonora Bobcat: cloud-free, substellar atmosphere models, spectra, photometry, evolution, and chemistry, Sonora Bobcat,  Zenodo, \dodoi{10.5281/zenodo.5063476}

\bibitem[{{Mawet} {et~al.}(2019){Mawet}, {Hirsch}, {Lee}, {Ruffio}, {Bottom}, {Fulton}, {Absil}, {Beichman}, {Bowler}, {Bryan}, {Choquet}, {Ciardi}, {Christiaens}, {Defr{\`e}re}, {Gomez Gonzalez}, {Howard}, {Huby}, {Isaacson}, {Jensen-Clem}, {Kosiarek}, {Marcy}, {Meshkat}, {Petigura}, {Reggiani}, {Ruane}, {Serabyn}, {Sinukoff}, {Wang}, {Weiss}, \& {Ygouf}}]{epseri_mawet_2019}
{Mawet}, D., {Hirsch}, L., {Lee}, E.~J., {et~al.} 2019, \aj, 157, 33, \dodoi{10.3847/1538-3881/aaef8a}

\bibitem[{{Nielsen} {et~al.}(2020){Nielsen}, {De Rosa}, {Wang}, {Sahlmann}, {Kalas}, {Duch{\^e}ne}, {Rameau}, {Marley}, {Saumon}, {Macintosh}, {Millar-Blanchaer}, {Nguyen}, {Ammons}, {Bailey}, {Barman}, {Bulger}, {Chilcote}, {Cotten}, {Doyon}, {Esposito}, {Fitzgerald}, {Follette}, {Gerard}, {Goodsell}, {Graham}, {Greenbaum}, {Hibon}, {Hung}, {Ingraham}, {Konopacky}, {Larkin}, {Maire}, {Marchis}, {Marois}, {Metchev}, {Oppenheimer}, {Palmer}, {Patience}, {Perrin}, {Poyneer}, {Pueyo}, {Rajan}, {Rantakyr{\"o}}, {Ruffio}, {Savransky}, {Schneider}, {Sivaramakrishnan}, {Song}, {Soummer}, {Thomas}, {Wallace}, {Ward-Duong}, {Wiktorowicz}, \& {Wolff}}]{betapic_nielsen_2020}
{Nielsen}, E.~L., {De Rosa}, R.~J., {Wang}, J.~J., {et~al.} 2020, \aj, 159, 71, \dodoi{10.3847/1538-3881/ab5b92}

\bibitem[{{Pathak} {et~al.}(2021){Pathak}, {Petit dit de la Roche}, {Kasper}, {Sterzik}, {Absil}, {Boehle}, {Feng}, {Ivanov}, {Janson}, {Jones}, {Kaufer}, {K{\"a}ufl}, {Maire}, {Meyer}, {Pantin}, {Siebenmorgen}, {van den Ancker}, \& {Viswanath}}]{epseri_near_pathak_2021}
{Pathak}, P., {Petit dit de la Roche}, D.~J.~M., {Kasper}, M., {et~al.} 2021, \aap, 652, A121, \dodoi{10.1051/0004-6361/202140529}

\bibitem[{Rains {et~al.}(2020)Rains, Ireland, White, Casagrande, \& Karovicova}]{stellar_radii_rains_2020}
Rains, A.~D., Ireland, M.~J., White, T.~R., Casagrande, L., \& Karovicova, I. 2020, Monthly Notices of the Royal Astronomical Society, 493, 2377, \dodoi{10.1093/mnras/staa282}

\bibitem[{{Reffert} \& {Quirrenbach}(2011)}]{HipparcosIAD_planets_Reffert_2011}
{Reffert}, S., \& {Quirrenbach}, A. 2011, \aap, 527, A140, \dodoi{10.1051/0004-6361/201015861}

\bibitem[{{Roettenbacher} {et~al.}(2022){Roettenbacher}, {Cabot}, {Fischer}, {Monnier}, {Henry}, {Harmon}, {Korhonen}, {Brewer}, {Llama}, {Petersburg}, {Zhao}, {Kraus}, {Le Bouquin}, {Anugu}, {Davies}, {Gardner}, {Lanthermann}, {Schaefer}, {Setterholm}, {Clark}, {Jorstad}, {Kuehn}, \& {Levine}}]{epseri_express_Roettenbacher_2022}
{Roettenbacher}, R.~M., {Cabot}, S. H.~C., {Fischer}, D.~A., {et~al.} 2022, \aj, 163, 19, \dodoi{10.3847/1538-3881/ac3235}

\bibitem[{Rohatgi(2025)}]{WebPlotDigitizer}
Rohatgi, A. 2025, WebPlotDigitizer, 5.2.
\newblock \url{https://automeris.io}

\bibitem[{Surjanovic {et~al.}(2023)Surjanovic, Biron-Lattes, Tiede, Syed, Campbell, \& Bouchard-C{\^o}t{\'e}}]{pigeons}
Surjanovic, N., Biron-Lattes, M., Tiede, P., {et~al.} 2023, arXiv:2308.09769

\bibitem[{Surjanovic {et~al.}(2022)Surjanovic, Syed, Bouchard-C\^{o}t\'{e}, \& Campbell}]{stabvarpartemp}
Surjanovic, N., Syed, S., Bouchard-C\^{o}t\'{e}, A., \& Campbell, T. 2022, 35, 565.
\newblock \url{https://proceedings.neurips.cc/paper_files/paper/2022/file/03cd3cf3f74d4f9ce5958de269960884-Paper-Conference.pdf}

\bibitem[{Syed {et~al.}(2021)Syed, Bouchard-Côté, Deligiannidis, \& Doucet}]{nonreversept}
Syed, S., Bouchard-Côté, A., Deligiannidis, G., \& Doucet, A. 2021, Journal of the Royal Statistical Society Series B: Statistical Methodology, 84, 321, \dodoi{10.1111/rssb.12464}

\bibitem[{Thompson {et~al.}(2022)Thompson, Marois, Do~Ó, Konopacky, Ruffio, Wang, Skemer, De~Rosa, \& Macintosh}]{hr8799_Thompson_2023}
Thompson, W., Marois, C., Do~Ó, C.~R., {et~al.} 2022, The Astronomical Journal, 165, 29, \dodoi{10.3847/1538-3881/aca1af}

\bibitem[{{Thompson} {et~al.}(2023){Thompson}, {Lawrence}, {Blakely}, {Marois}, {Wang}, {Giordano}, {Brandt}, {Johnstone}, {Ruffio}, {Ammons}, {Crotts}, {Do {\'O}}, {Gonzales}, \& {Rice}}]{thompson_octofitter_2023}
{Thompson}, W., {Lawrence}, J., {Blakely}, D., {et~al.} 2023, \aj, 166, 164, \dodoi{10.3847/1538-3881/acf5cc}

\bibitem[{{Trifonov} {et~al.}(2020){Trifonov}, {Tal-Or}, {Zechmeister}, {Kaminski}, {Zucker}, \& {Mazeh}}]{rvs_harps_Trifonov_2020}
{Trifonov}, T., {Tal-Or}, L., {Zechmeister}, M., {et~al.} 2020, \aap, 636, A74, \dodoi{10.1051/0004-6361/201936686}

\bibitem[{{van Leeuwen}(2007)}]{hipparcos_van_Leeuwen_2007}
{van Leeuwen}, F. 2007, {Hipparcos, the New Reduction of the Raw Data}, Vol. 350, \dodoi{10.1007/978-1-4020-6342-8}

\bibitem[{{van Leeuwen} {et~al.}(1997){van Leeuwen}, {Evans}, {Grenon}, {Grossmann}, {Mignard}, \& {Perryman}}]{Hipparcos_van_leeuwen_1997}
{van Leeuwen}, F., {Evans}, D.~W., {Grenon}, M., {et~al.} 1997, \aap, 323, L61

\bibitem[{Vehtari {et~al.}(2021)Vehtari, Gelman, Simpson, Carpenter, \& B{\"u}rkner}]{rhat_vehtari_2021}
Vehtari, A., Gelman, A., Simpson, D., Carpenter, B., \& B{\"u}rkner, P.-C. 2021, Bayesian Analysis, 16, 667 , \dodoi{10.1214/20-BA1221}

\bibitem[{{Zechmeister} {et~al.}(2013){Zechmeister}, {K{\"u}rster}, {Endl}, {Lo Curto}, {Hartman}, {Nilsson}, {Henning}, {Hatzes}, \& {Cochran}}]{rvs_ces_Zechmeister_2013}
{Zechmeister}, M., {K{\"u}rster}, M., {Endl}, M., {et~al.} 2013, \aap, 552, A78, \dodoi{10.1051/0004-6361/201116551}

\end{thebibliography}
\bibliographystyle{aasjournal}

\appendix

\section{Additional Tables}

We now present summary statistics for each variable of our models. Table \ref{tab:hdi} presents the 75\% HDI, while Table \ref{tab:ci} presents the 68\% CI.

The residuals we digitized from Figure 7 of \citet{fgs_eps_eri_benedict_2006} are also presented in Tables \ref{tab:fgs-eta} and \ref{tab:fgs-xi}. Based on that figure, we interpret the $\eta$ coordinate as a \emph{negative} change in declination, while $\xi$ is a \emph{negative} change in right ascension, times the cosine of declination.

\startlongtable
\begin{deluxetable*}{llcccccc}
\centerwidetable
\tablecaption{Posterior 75\% Highest Density Intervals.\label{tab:hdi}}
\tabletypesize{\tiny}
\tablehead{\colhead{Parameter} & \colhead{RV} & \colhead{HIP, FGS, DR2, DR3} & \colhead{RV, FGS} & \colhead{RV, Hip. IAD} & \colhead{RV, HGCA} & \colhead{RV, HIP, FGS, DR3} & \colhead{RV, HIP, FGS, DR2, DR3}}
\startdata
$m_{A}$ & $[0.80, 0.84]$ & $[0.80, 0.84]$ & $[0.79, 0.84]$ & $[0.79, 0.84]$ & $[0.80, 0.84]$ & $[0.80, 0.84]$ & $[0.80, 0.84]$ \\
$m_{b}$ & $[0.56, 1.58]$ & $[0.78, 1.44]$ & $[0.59, 0.84]$ & $[0.56, 1.09]$ & $[0.64, 0.95]$ & $[0.68, 0.95]$ & $[0.87, 1.10]$ \\
$P^!$ & $[7.26, 7.46]$ & $[6.95, 7.53]$ & $[7.28, 7.47]$ & $[7.25, 7.45]$ & $[7.25, 7.44]$ & $[7.24, 7.42]$ & $[7.24, 7.41]$ \\
$a$ & $[3.50, 3.59]$ & $[3.41, 3.61]$ & $[3.49, 3.59]$ & $[3.50, 3.59]$ & $[3.49, 3.58]$ & $[3.49, 3.58]$ & $[3.49, 3.58]$ \\
$e$ & $[0.00, 0.11]$ & $[0.00, 0.44]$ & $[0.00, 0.11]$ & $[0.00, 0.11]$ & $[0.00, 0.11]$ & $[0.00, 0.10]$ & $[0.00, 0.10]$ \\
$i$ & $[24.12, 154.78]$ & $[28.04, 76.89]$ & $[36.56, 83.88]$ & $[19.34, 120.36]$ & $[42.15, 135.53]$ & $[36.65, 61.20]$ & $[33.65, 46.55]$ \\
$t_{p^!}$ & $[57389.58, 59337.73]$ & $[58277.49, 60055.43]$ & $[57399.66, 58760.38]$ & $[48919.03, 50005.17]$ & $[57388.83, 59486.04]$ & $[57388.52, 59272.43]$ & $[57389.04, 59521.86]$ \\
$\Omega$ & $[1.75, 263.99]$ & $[10.51, 225.35]$ & $[156.33, 192.81]$ & $[101.05, 238.39]$ & $[179.67, 216.03]$ & $[173.03, 198.48]$ & $[176.79, 196.43]$ \\
$\tau$ & $[0.00, 0.73]$ & $[0.37, 1.00]$ & $[0.00, 0.51]$ & $[0.21, 0.61]$ & $[0.00, 0.78]$ & $[0.00, 0.71]$ & $[0.00, 0.80]$ \\
$\omega$ & $[193.23, 359.82]$ & $[85.09, 347.48]$ & $[75.46, 358.92]$ & $[208.77, 360.00]$ & $[183.90, 359.98]$ & $[169.73, 359.81]$ & $[198.06, 359.98]$ \\
$B$ & $[51.36, 61.72]$ & - & $[50.88, 61.58]$ & $[50.90, 61.84]$ & $[51.59, 62.54]$ & $[50.95, 61.99]$ & $[50.66, 61.90]$ \\
$C$ & $[0.00, 0.14]$ & - & $[0.00, 0.14]$ & $[0.00, 0.14]$ & $[0.00, 0.13]$ & $[0.00, 0.14]$ & $[0.00, 0.14]$ \\
$L$ & $[5.82, 8.68]$ & - & $[6.01, 8.90]$ & $[5.85, 8.78]$ & $[5.94, 8.81]$ & $[5.86, 8.75]$ & $[5.90, 8.78]$ \\
$P_{rot}$ & $[10.42, 11.62]$ & - & $[10.36, 11.62]$ & $[10.44, 11.65]$ & $[10.55, 11.74]$ & $[10.46, 11.67]$ & $[10.53, 11.71]$ \\
$\sigma_{APF}$ & $[0.75, 1.80]$ & - & $[0.72, 1.77]$ & $[0.80, 1.86]$ & $[0.81, 1.87]$ & $[0.77, 1.84]$ & $[0.76, 1.81]$ \\
$\sigma_{EXPRES}$ & $[0.76, 0.90]$ & - & $[0.74, 0.90]$ & $[0.76, 0.91]$ & $[0.76, 0.90]$ & $[0.76, 0.91]$ & $[0.75, 0.90]$ \\
$\sigma_{CFHT}$ & $[0.10, 6.25]$ & - & $[0.11, 6.70]$ & $[0.10, 6.39]$ & $[0.15, 6.34]$ & $[0.10, 6.19]$ & $[0.10, 6.10]$ \\
$\sigma_{CHIRON}$ & $[0.10, 1.31]$ & - & $[0.11, 1.28]$ & $[0.10, 1.28]$ & $[0.10, 1.39]$ & $[0.10, 1.31]$ & $[0.10, 1.28]$ \\
$\sigma_{HIRES}$ & $[0.54, 1.44]$ & - & $[0.67, 1.47]$ & $[0.54, 1.44]$ & $[0.52, 1.48]$ & $[0.58, 1.47]$ & $[0.59, 1.47]$ \\
$\sigma_{Lick_1}$ & $[7.19, 13.19]$ & - & $[6.81, 12.64]$ & $[6.59, 12.68]$ & $[7.13, 12.89]$ & $[6.58, 12.76]$ & $[6.41, 12.56]$ \\
$\sigma_{Lick_2}$ & $[0.10, 5.49]$ & - & $[0.13, 5.34]$ & $[0.10, 5.44]$ & $[0.10, 5.72]$ & $[0.10, 5.45]$ & $[0.10, 5.57]$ \\
$\sigma_{Lick_3}$ & $[0.10, 3.48]$ & - & $[0.12, 3.29]$ & $[0.10, 3.52]$ & $[0.13, 3.50]$ & $[0.11, 3.50]$ & $[0.10, 3.48]$ \\
$\sigma_{Lick_4}$ & $[2.93, 5.02]$ & - & $[2.93, 4.97]$ & $[2.93, 5.09]$ & $[2.78, 4.92]$ & $[2.86, 5.04]$ & $[2.93, 5.06]$ \\
$\sigma_{HARPS}$ & $[0.10, 0.17]$ & - & $[0.10, 0.15]$ & $[0.10, 0.13]$ & $[0.10, 0.14]$ & $[0.10, 0.13]$ & $[0.10, 0.15]$ \\
$\sigma_{CESlc}$ & $[0.10, 2.94]$ & - & $[0.12, 3.12]$ & $[0.10, 2.85]$ & $[0.11, 2.88]$ & $[0.10, 2.93]$ & $[0.10, 2.89]$ \\
$\sigma_{CESvlc}$ & $[0.11, 1.55]$ & - & $[0.10, 1.58]$ & $[0.10, 1.47]$ & $[0.10, 1.48]$ & $[0.10, 1.53]$ & $[0.10, 1.51]$ \\
$m_{APF}$ & $[-0.00, -0.00]$ & - & $[-0.00, -0.00]$ & $[-0.00, -0.00]$ & $[-0.00, -0.00]$ & $[-0.00, -0.00]$ & $[-0.00, -0.00]$ \\
$m_{EXPRES}$ & $[-0.02, 0.01]$ & - & $[-0.01, 0.01]$ & $[-0.02, 0.01]$ & $[-0.01, 0.01]$ & $[-0.01, 0.01]$ & $[-0.01, 0.01]$ \\
$m_{CFHT}$ & $[0.00, 0.02]$ & - & $[0.00, 0.02]$ & $[0.00, 0.02]$ & $[0.00, 0.01]$ & $[0.00, 0.01]$ & $[0.00, 0.01]$ \\
$m_{CHIRON}$ & $[-0.05, 0.07]$ & - & $[-0.04, 0.06]$ & $[-0.05, 0.07]$ & $[-0.04, 0.08]$ & $[-0.05, 0.07]$ & $[-0.04, 0.08]$ \\
$m_{HIRES}$ & $[-0.00, 0.00]$ & - & $[-0.00, 0.00]$ & $[-0.00, 0.00]$ & $[-0.00, 0.00]$ & $[-0.00, 0.00]$ & $[-0.00, 0.00]$ \\
$m_{Lick_1}$ & $[0.00, 0.01]$ & - & $[0.00, 0.01]$ & $[0.00, 0.01]$ & $[0.00, 0.01]$ & $[0.00, 0.01]$ & $[0.00, 0.01]$ \\
$m_{Lick_2}$ & $[-0.00, 0.02]$ & - & $[-0.00, 0.02]$ & $[-0.00, 0.02]$ & $[-0.01, 0.01]$ & $[-0.00, 0.02]$ & $[-0.00, 0.02]$ \\
$m_{Lick_3}$ & $[-0.00, 0.01]$ & - & $[-0.00, 0.01]$ & $[-0.00, 0.01]$ & $[-0.00, 0.01]$ & $[-0.00, 0.01]$ & $[-0.00, 0.01]$ \\
$m_{Lick_4}$ & $[0.00, 0.00]$ & - & $[0.00, 0.00]$ & $[0.00, 0.00]$ & $[0.00, 0.00]$ & $[0.00, 0.00]$ & $[0.00, 0.00]$ \\
$m_{HARPSpre}$ & $[-0.01, -0.00]$ & - & $[-0.01, 0.00]$ & $[-0.01, 0.00]$ & $[-0.01, -0.00]$ & $[-0.01, -0.00]$ & $[-0.01, -0.00]$ \\
$m_{HARPSpost}$ & $[-0.01, 0.00]$ & - & $[-0.01, 0.00]$ & $[-0.01, 0.00]$ & $[-0.01, 0.00]$ & $[-0.01, 0.00]$ & $[-0.01, 0.00]$ \\
$m_{CESlc}$ & $[-0.01, 0.00]$ & - & $[-0.01, 0.00]$ & $[-0.01, 0.00]$ & $[-0.01, 0.00]$ & $[-0.01, 0.00]$ & $[-0.01, 0.00]$ \\
$m_{CESvlc}$ & $[-0.00, 0.00]$ & - & $[-0.00, 0.00]$ & $[-0.00, 0.00]$ & $[-0.00, 0.01]$ & $[-0.00, 0.01]$ & $[-0.00, 0.01]$ \\
$\gamma_{FGS\_\delta}$ & - & $[-0.00, 0.00]$ & $[-0.00, -0.00]$ & - & - & $[-0.00, -0.00]$ & $[-0.00, 0.00]$ \\
$\gamma_{FGS\_\alpha✱}$ & - & $[0.00, 0.00]$ & $[0.00, 0.00]$ & - & - & $[0.00, 0.00]$ & $[0.00, 0.00]$ \\
$\omega_{x}$ & - & $[-0.13, -0.01]$ & - & - & - & - & $[-0.14, -0.02]$ \\
$\omega_{y}$ & - & $[-0.10, -0.00]$ & - & - & - & - & $[-0.10, -0.00]$ \\
$\omega_{z}$ & - & $[-0.06, 0.03]$ & - & - & - & - & $[-0.05, 0.03]$ \\
$\varpi$ & $[310.43, 310.73]$ & $[310.44, 310.75]$ & $[310.44, 310.72]$ & $[310.68, 311.07]$ & $[310.43, 310.72]$ & $[310.42, 310.73]$ & $[310.42, 310.73]$ \\
$\mu_{\delta}$ & $[-8.98, 47.93]$ & $[19.98, 20.06]$ & $[-2.85, 51.18]$ & $[18.98, 20.08]$ & $[19.99, 20.03]$ & $[19.97, 20.02]$ & $[19.98, 20.02]$ \\
$\mu_{\alpha✱}$ & $[-1003.81, -945.58]$ & $[-975.03, -974.96]$ & $[-1000.94, -946.41]$ & $[-975.88, -974.43]$ & $[-975.05, -974.95]$ & $[-975.04, -974.98]$ & $[-974.96, -974.91]$ \\
$\gamma_{APF}$ & $[-1.28, 0.97]$ & - & $[-1.22, 0.79]$ & $[-2.99, -0.78]$ & $[-1.23, 0.90]$ & $[-1.29, 0.91]$ & $[-1.36, 0.88]$ \\
$\gamma_{EXPRES}$ & $[-9.04, -4.16]$ & - & $[-9.34, -5.02]$ & $[-10.81, -6.10]$ & $[-8.98, -4.39]$ & $[-9.36, -4.63]$ & $[-9.38, -4.54]$ \\
$\gamma_{CFHT}$ & $[-0.38, 7.07]$ & - & $[-0.66, 6.51]$ & $[-2.43, 5.11]$ & $[-0.79, 6.61]$ & $[-0.63, 6.83]$ & $[-0.72, 6.70]$ \\
$\gamma_{CHIRON}$ & $[-15.84, -10.83]$ & - & $[-15.01, -10.43]$ & $[-17.56, -12.56]$ & $[-15.92, -10.95]$ & $[-15.89, -10.75]$ & $[-15.82, -10.82]$ \\
$\gamma_{HIRES}$ & $[3.18, 5.20]$ & - & $[3.32, 5.37]$ & $[1.51, 3.47]$ & $[3.44, 5.44]$ & $[3.19, 5.22]$ & $[3.33, 5.31]$ \\
$\gamma_{Lick_1}$ & $[6.55, 11.62]$ & - & $[5.80, 11.78]$ & $[4.56, 9.79]$ & $[6.53, 11.82]$ & $[6.36, 11.62]$ & $[6.22, 11.28]$ \\
$\gamma_{Lick_2}$ & $[6.19, 13.47]$ & - & $[6.23, 13.78]$ & $[4.10, 11.79]$ & $[6.23, 13.69]$ & $[5.99, 13.59]$ & $[6.28, 13.58]$ \\
$\gamma_{Lick_3}$ & $[9.13, 14.18]$ & - & $[9.04, 14.04]$ & $[7.26, 12.24]$ & $[8.88, 13.91]$ & $[8.74, 13.67]$ & $[8.59, 13.45]$ \\
$\gamma_{Lick_4}$ & $[-2.56, 0.03]$ & - & $[-2.24, 0.41]$ & $[-4.23, -1.51]$ & $[-2.15, 0.36]$ & $[-2.46, 0.17]$ & $[-2.50, 0.18]$ \\
$\gamma_{HARPSpre}$ & $[-12.05, -7.27]$ & - & $[-12.02, -7.30]$ & $[-13.87, -9.19]$ & $[-11.83, -7.31]$ & $[-11.55, -6.97]$ & $[-11.65, -6.98]$ \\
$\gamma_{HARPSpost}$ & $[-6.27, -1.11]$ & - & $[-6.68, -1.91]$ & $[-7.94, -2.82]$ & $[-6.45, -1.33]$ & $[-6.42, -1.34]$ & $[-6.39, -1.12]$ \\
$\gamma_{CESlc}$ & $[7.21, 12.46]$ & - & $[7.43, 12.22]$ & $[5.36, 10.62]$ & $[6.83, 12.00]$ & $[6.94, 12.28]$ & $[7.23, 12.53]$ \\
$\gamma_{CESvlc}$ & $[2.19, 6.89]$ & - & $[2.33, 6.50]$ & $[0.40, 5.09]$ & $[2.19, 6.97]$ & $[2.28, 6.96]$ & $[2.02, 6.74]$ \\
$\gamma_{FGS\_dec}$ & - & $[1.03, 2.21]$ & $[1.69, 2.06]$ & - & - & $[1.58, 1.94]$ & $[1.40, 1.76]$ \\
$\gamma_{FGS\_ra}$ & - & $[-0.82, -0.05]$ & $[-0.60, -0.14]$ & - & - & $[-0.67, -0.30]$ & $[-0.63, -0.26]$ \\
$\sigma_{FGS}$ & - & $[0.52, 0.62]$ & $[0.52, 0.63]$ & - & - & $[0.52, 0.62]$ & $[0.52, 0.62]$ \\
\enddata
\end{deluxetable*}

\startlongtable
\begin{deluxetable*}{llcccccc}
\centerwidetable
\tablecaption{Posterior 68\% Credible Intervals. \label{tab:ci}}
\tabletypesize{\tiny}
\tablehead{\colhead{Parameter} & \colhead{RV} & \colhead{HIP, FGS, DR2, DR3} & \colhead{RV, FGS} & \colhead{RV, Hip. IAD} & \colhead{RV, HGCA} & \colhead{RV, HIP, FGS, DR3} & \colhead{RV, HIP, FGS, DR2, DR3}}
\startdata
$m_{A}$ & $0.820^{+0.020}_{-0.021}$ & $0.820 \pm 0.020$ & $0.817^{+0.024}_{-0.020}$ & $0.820 \pm 0.020$ & $0.820^{+0.020}_{-0.019}$ & $0.820 \pm 0.020$ & $0.820 \pm 0.020$ \\
$m_{b}$ & $1.0^{+1.2}_{-0.3}$ & $1.1^{+0.5}_{-0.2}$ & $0.72^{+0.15}_{-0.08}$ & $0.8^{+0.5}_{-0.2}$ & $0.83^{+0.16}_{-0.14}$ & $0.83^{+0.12}_{-0.11}$ & $1.00 \pm 0.10$ \\
$P^!$ & $7.36 \pm 0.09$ & $7.26^{+0.22}_{-0.27}$ & $7.36^{+0.08}_{-0.09}$ & $7.37 \pm 0.09$ & $7.34^{+0.09}_{-0.07}$ & $7.33 \pm 0.08$ & $7.33^{+0.08}_{-0.07}$ \\
$a$ & $3.54 \pm 0.04$ & $3.51^{+0.08}_{-0.09}$ & $3.54^{+0.05}_{-0.04}$ & $3.54 \pm 0.04$ & $3.54^{+0.04}_{-0.03}$ & $3.53 \pm 0.04$ & $3.53 \pm 0.04$ \\
$e$ & $0.07^{+0.06}_{-0.05}$ & $0.2^{+0.3}_{-0.2}$ & $0.06^{+0.07}_{-0.05}$ & $0.07^{+0.06}_{-0.05}$ & $0.06^{+0.07}_{-0.04}$ & $0.06^{+0.06}_{-0.04}$ & $0.06^{+0.06}_{-0.04}$ \\
$i$ & $90^{+60}_{-62}$ & $54^{+19}_{-23}$ & $63^{+30}_{-17}$ & $78^{+53}_{-43}$ & $76^{+59}_{-31}$ & $50^{+14}_{-9}$ & $40^{+6}_{-5}$ \\
$t_{p^!}$ & $57969^{+1692}_{-409}$ & $59185^{+592}_{-1370}$ & $58030^{+1001}_{-419}$ & $49543^{+420}_{-470}$ & $58131^{+1609}_{-523}$ & $57998^{+1680}_{-390}$ & $58017^{+1740}_{-447}$ \\
$\Omega$ & $176 \pm 119$ & $182^{+54}_{-157}$ & $174^{+15}_{-17}$ & $168^{+61}_{-58}$ & $198 \pm 16$ & $186 \pm 11$ & $186^{+8}_{-9}$ \\
$\tau$ & $0.2^{+0.6}_{-0.2}$ & $0.7^{+0.2}_{-0.5}$ & $0.2^{+0.4}_{-0.2}$ & $0.44^{+0.16}_{-0.17}$ & $0.3^{+0.6}_{-0.2}$ & $0.2^{+0.6}_{-0.1}$ & $0.2^{+0.7}_{-0.2}$ \\
$\omega$ & $279^{+49}_{-183}$ & $179^{+121}_{-120}$ & $206^{+121}_{-191}$ & $282^{+52}_{-154}$ & $268^{+64}_{-134}$ & $275^{+56}_{-190}$ & $268^{+62}_{-105}$ \\
$B$ & $57^{+5}_{-4}$ & - & $57 \pm 5$ & $57 \pm 5$ & $57^{+5}_{-4}$ & $57^{+5}_{-4}$ & $57 \pm 5$ \\
$C$ & $0.06^{+0.13}_{-0.05}$ & - & $0.07^{+0.11}_{-0.05}$ & $0.07^{+0.12}_{-0.05}$ & $0.06^{+0.11}_{-0.05}$ & $0.07^{+0.12}_{-0.05}$ & $0.07^{+0.12}_{-0.05}$ \\
$L$ & $7.3^{+1.4}_{-1.1}$ & - & $7.2^{+1.5}_{-1.1}$ & $7.3^{+1.4}_{-1.2}$ & $7.4^{+1.5}_{-1.2}$ & $7.4^{+1.4}_{-1.1}$ & $7.4^{+1.5}_{-1.1}$ \\
$P_{rot}$ & $11.1 \pm 0.5$ & - & $11.0^{+0.5}_{-0.6}$ & $11.1 \pm 0.5$ & $11.1 \pm 0.5$ & $11.1 \pm 0.5$ & $11.1 \pm 0.5$ \\
$\sigma_{APF}$ & $1.3^{+0.4}_{-0.5}$ & - & $1.3^{+0.4}_{-0.5}$ & $1.3^{+0.4}_{-0.5}$ & $1.3^{+0.4}_{-0.5}$ & $1.3^{+0.4}_{-0.5}$ & $1.3^{+0.4}_{-0.5}$ \\
$\sigma_{EXPRES}$ & $0.83^{+0.07}_{-0.06}$ & - & $0.84^{+0.08}_{-0.06}$ & $0.83^{+0.07}_{-0.06}$ & $0.83 \pm 0.06$ & $0.83^{+0.07}_{-0.06}$ & $0.83^{+0.07}_{-0.06}$ \\
$\sigma_{CFHT}$ & $4^{+4}_{-2}$ & - & $4^{+4}_{-3}$ & $4^{+4}_{-3}$ & $4^{+4}_{-2}$ & $4^{+4}_{-3}$ & $4^{+4}_{-2}$ \\
$\sigma_{CHIRON}$ & $0.8^{+0.8}_{-0.5}$ & - & $0.8^{+0.8}_{-0.5}$ & $0.8^{+0.8}_{-0.5}$ & $0.8^{+0.8}_{-0.6}$ & $0.8^{+0.8}_{-0.5}$ & $0.8^{+0.8}_{-0.5}$ \\
$\sigma_{HIRES}$ & $1.0 \pm 0.4$ & - & $1.1^{+0.4}_{-0.3}$ & $1.0 \pm 0.4$ & $1.0 \pm 0.4$ & $1.0 \pm 0.4$ & $1.0 \pm 0.4$ \\
$\sigma_{Lick_1}$ & $9.8^{+2.7}_{-2.5}$ & - & $10.4^{+2.4}_{-2.5}$ & $9.9 \pm 2.6$ & $9.6^{+2.5}_{-2.6}$ & $9.6^{+2.6}_{-2.7}$ & $9.6^{+2.6}_{-2.8}$ \\
$\sigma_{Lick_2}$ & $3^{+4}_{-2}$ & - & $3^{+4}_{-2}$ & $3^{+4}_{-2}$ & $3^{+4}_{-2}$ & $3^{+4}_{-2}$ & $3^{+4}_{-2}$ \\
$\sigma_{Lick_3}$ & $2.2^{+1.9}_{-1.4}$ & - & $2.1^{+2.0}_{-1.5}$ & $2.2^{+2.0}_{-1.4}$ & $2.2^{+2.0}_{-1.4}$ & $2.2^{+2.0}_{-1.4}$ & $2.2^{+2.0}_{-1.4}$ \\
$\sigma_{Lick_4}$ & $3.9 \pm 0.9$ & - & $3.9 \pm 0.9$ & $3.9 \pm 0.9$ & $3.9^{+1.0}_{-0.9}$ & $4.0^{+1.0}_{-0.9}$ & $4.0 \pm 0.9$ \\
$\sigma_{HARPS}$ & $0.118^{+0.026}_{-0.013}$ & - & $0.117^{+0.024}_{-0.012}$ & $0.118^{+0.026}_{-0.013}$ & $0.117^{+0.025}_{-0.013}$ & $0.118^{+0.025}_{-0.013}$ & $0.117^{+0.025}_{-0.013}$ \\
$\sigma_{CESlc}$ & $1.8^{+1.7}_{-1.2}$ & - & $2.0^{+1.5}_{-1.4}$ & $1.7^{+1.7}_{-1.1}$ & $1.6^{+1.9}_{-1.0}$ & $1.8^{+1.8}_{-1.2}$ & $1.8^{+1.7}_{-1.2}$ \\
$\sigma_{CESvlc}$ & $0.9^{+1.0}_{-0.6}$ & - & $0.9^{+0.9}_{-0.6}$ & $0.9^{+0.9}_{-0.6}$ & $0.9^{+0.9}_{-0.6}$ & $0.9^{+0.9}_{-0.6}$ & $0.9^{+0.9}_{-0.6}$ \\
$m_{APF}$ & $-0.0023^{+0.0016}_{-0.0018}$ & - & $-0.0024 \pm 0.0018$ & $-0.0023^{+0.0017}_{-0.0018}$ & $-0.0022 \pm 0.0016$ & $-0.0024 \pm 0.0017$ & $-0.0023 \pm 0.0017$ \\
$m_{EXPRES}$ & $-0.002 \pm 0.011$ & - & $-0.001^{+0.010}_{-0.011}$ & $-0.002 \pm 0.012$ & $-0.002 \pm 0.011$ & $-0.002 \pm 0.011$ & $-0.002 \pm 0.011$ \\
$m_{CFHT}$ & $0.010 \pm 0.005$ & - & $0.010 \pm 0.005$ & $0.010 \pm 0.005$ & $0.010 \pm 0.005$ & $0.009 \pm 0.005$ & $0.009^{+0.005}_{-0.004}$ \\
$m_{CHIRON}$ & $0.01 \pm 0.05$ & - & $0.01^{+0.04}_{-0.05}$ & $0.01 \pm 0.05$ & $0.01 \pm 0.05$ & $0.01 \pm 0.05$ & $0.01 \pm 0.05$ \\
$m_{HIRES}$ & $-0.0006 \pm 0.0007$ & - & $-0.0007^{+0.0008}_{-0.0006}$ & $-0.0006 \pm 0.0007$ & $-0.0007^{+0.0007}_{-0.0006}$ & $-0.0007 \pm 0.0007$ & $-0.0007 \pm 0.0007$ \\
$m_{Lick_1}$ & $0.0040^{+0.0028}_{-0.0027}$ & - & $0.004 \pm 0.003$ & $0.0040^{+0.0029}_{-0.0028}$ & $0.0038^{+0.0027}_{-0.0028}$ & $0.0037^{+0.0028}_{-0.0027}$ & $0.0036 \pm 0.0027$ \\
$m_{Lick_2}$ & $0.005 \pm 0.009$ & - & $0.006^{+0.008}_{-0.009}$ & $0.005 \pm 0.009$ & $0.006^{+0.009}_{-0.008}$ & $0.006 \pm 0.009$ & $0.006 \pm 0.008$ \\
$m_{Lick_3}$ & $0.003 \pm 0.005$ & - & $0.002^{+0.006}_{-0.005}$ & $0.003 \pm 0.005$ & $0.003 \pm 0.005$ & $0.002 \pm 0.005$ & $0.002 \pm 0.005$ \\
$m_{Lick_4}$ & $0.0030^{+0.0011}_{-0.0012}$ & - & $0.0030^{+0.0010}_{-0.0012}$ & $0.0030 \pm 0.0012$ & $0.0030^{+0.0012}_{-0.0013}$ & $0.0029 \pm 0.0012$ & $0.0029 \pm 0.0012$ \\
$m_{HARPSpre}$ & $-0.005^{+0.004}_{-0.005}$ & - & $-0.004^{+0.004}_{-0.005}$ & $-0.005 \pm 0.004$ & $-0.005 \pm 0.004$ & $-0.005 \pm 0.004$ & $-0.005 \pm 0.004$ \\
$m_{HARPSpost}$ & $-0.002 \pm 0.004$ & - & $-0.003 \pm 0.003$ & $-0.002 \pm 0.004$ & $-0.003 \pm 0.004$ & $-0.003 \pm 0.004$ & $-0.003 \pm 0.004$ \\
$m_{CESlc}$ & $-0.004^{+0.004}_{-0.005}$ & - & $-0.004^{+0.004}_{-0.005}$ & $-0.004 \pm 0.005$ & $-0.003 \pm 0.004$ & $-0.003^{+0.005}_{-0.004}$ & $-0.003 \pm 0.004$ \\
$m_{CESvlc}$ & $0.001 \pm 0.003$ & - & $0.001 \pm 0.003$ & $0.001 \pm 0.003$ & $0.001 \pm 0.003$ & $0.001 \pm 0.003$ & $0.001 \pm 0.003$ \\
$\gamma_{FGS\_\delta}$ & - & $0.0003^{+0.0007}_{-0.0008}$ & $-0.0004 \pm 0.0004$ & - & - & $-0.0004 \pm 0.0003$ & $-0.0003 \pm 0.0004$ \\
$\gamma_{FGS\_\alpha✱}$ & - & $0.0036 \pm 0.0008$ & $0.0025^{+0.0008}_{-0.0009}$ & - & - & $0.0031^{+0.0005}_{-0.0006}$ & $0.0037 \pm 0.0004$ \\
$\omega_{x}$ & - & $-0.07 \pm 0.05$ & - & - & - & - & $-0.08 \pm 0.05$ \\
$\omega_{y}$ & - & $-0.05 \pm 0.04$ & - & - & - & - & $-0.05 \pm 0.04$ \\
$\omega_{z}$ & - & $-0.02 \pm 0.04$ & - & - & - & - & $-0.02 \pm 0.04$ \\
$\varpi$ & $310.58 \pm 0.13$ & $310.58 \pm 0.14$ & $310.58 \pm 0.12$ & $310.87 \pm 0.17$ & $310.58^{+0.12}_{-0.13}$ & $310.58^{+0.14}_{-0.13}$ & $310.58 \pm 0.13$ \\
$\mu_{\delta}$ & $21^{+25}_{-24}$ & $20.02 \pm 0.04$ & $20^{+25}_{-23}$ & $19.6^{+0.4}_{-0.5}$ & $20.010^{+0.021}_{-0.022}$ & $19.996 \pm 0.018$ & $19.997^{+0.018}_{-0.017}$ \\
$\mu_{\alpha✱}$ & $-974^{+25}_{-26}$ & $-975.00 \pm 0.03$ & $-973^{+23}_{-27}$ & $-975.2 \pm 0.6$ & $-974.99^{+0.04}_{-0.05}$ & $-975.013^{+0.026}_{-0.027}$ & $-974.939^{+0.023}_{-0.022}$ \\
$\gamma_{APF}$ & $-0.2 \pm 0.9$ & - & $-0.2 \pm 0.9$ & $-2.0^{+1.0}_{-0.9}$ & $-0.2 \pm 0.9$ & $-0.2 \pm 1.0$ & $-0.3 \pm 1.0$ \\
$\gamma_{EXPRES}$ & $-6.8 \pm 2.1$ & - & $-7.2^{+1.9}_{-1.7}$ & $-8.5 \pm 2.1$ & $-6.8^{+2.0}_{-2.1}$ & $-7.0^{+2.1}_{-2.0}$ & $-6.9 \pm 2.1$ \\
$\gamma_{CFHT}$ & $3 \pm 3$ & - & $3 \pm 3$ & $2 \pm 3$ & $3 \pm 3$ & $3 \pm 3$ & $3 \pm 3$ \\
$\gamma_{CHIRON}$ & $-13.4 \pm 2.2$ & - & $-12.9 \pm 2.0$ & $-15.0^{+2.2}_{-2.1}$ & $-13.5 \pm 2.2$ & $-13.3 \pm 2.2$ & $-13.4^{+2.1}_{-2.3}$ \\
$\gamma_{HIRES}$ & $4.3 \pm 0.9$ & - & $4.3^{+0.8}_{-1.0}$ & $2.5^{+0.8}_{-0.9}$ & $4.3 \pm 0.9$ & $4.2 \pm 0.9$ & $4.2^{+0.9}_{-0.8}$ \\
$\gamma_{Lick_1}$ & $9.0 \pm 2.2$ & - & $9.1^{+2.4}_{-2.8}$ & $7.1 \pm 2.3$ & $8.9^{+2.4}_{-2.3}$ & $8.9 \pm 2.3$ & $8.7^{+2.3}_{-2.1}$ \\
$\gamma_{Lick_2}$ & $10 \pm 3$ & - & $10^{+4}_{-3}$ & $8 \pm 3$ & $10^{+4}_{-3}$ & $10 \pm 3$ & $10 \pm 3$ \\
$\gamma_{Lick_3}$ & $11.5^{+2.1}_{-2.2}$ & - & $11.4^{+2.3}_{-2.2}$ & $9.8 \pm 2.2$ & $11.2^{+2.4}_{-2.0}$ & $11.2^{+2.2}_{-2.1}$ & $11.0 \pm 2.1$ \\
$\gamma_{Lick_4}$ & $-1.1 \pm 1.1$ & - & $-1.0^{+0.9}_{-1.3}$ & $-2.9 \pm 1.2$ & $-1.0 \pm 1.1$ & $-1.1^{+1.1}_{-1.2}$ & $-1.0 \pm 1.2$ \\
$\gamma_{HARPSpre}$ & $-9.7^{+2.1}_{-2.0}$ & - & $-9.9^{+2.4}_{-1.9}$ & $-11.4^{+2.0}_{-2.1}$ & $-9.5^{+2.1}_{-1.9}$ & $-9.4^{+2.0}_{-1.9}$ & $-9.3^{+2.1}_{-2.0}$ \\
$\gamma_{HARPSpost}$ & $-3.6^{+2.2}_{-2.4}$ & - & $-4.1 \pm 2.0$ & $-5.3^{+2.3}_{-2.2}$ & $-3.6 \pm 2.2$ & $-3.8 \pm 2.2$ & $-3.7 \pm 2.3$ \\
$\gamma_{CESlc}$ & $9.8^{+2.4}_{-2.2}$ & - & $9.7^{+2.4}_{-2.0}$ & $8.1 \pm 2.3$ & $9.8^{+2.2}_{-2.4}$ & $9.7^{+2.3}_{-2.4}$ & $9.7 \pm 2.3$ \\
$\gamma_{CESvlc}$ & $4.6 \pm 2.1$ & - & $4.8^{+1.6}_{-2.0}$ & $2.8^{+2.0}_{-2.1}$ & $4.5 \pm 2.1$ & $4.5^{+2.0}_{-2.1}$ & $4.4^{+2.1}_{-2.0}$ \\
$\gamma_{FGS\_dec}$ & - & $1.6^{+0.4}_{-0.8}$ & $1.89^{+0.14}_{-0.20}$ & - & - & $1.75^{+0.15}_{-0.16}$ & $1.59^{+0.15}_{-0.16}$ \\
$\gamma_{FGS\_ra}$ & - & $-0.5^{+0.4}_{-0.3}$ & $-0.37 \pm 0.20$ & - & - & $-0.49^{+0.17}_{-0.16}$ & $-0.45 \pm 0.16$ \\
$\sigma_{FGS}$ & - & $0.57^{+0.05}_{-0.04}$ & $0.58 \pm 0.05$ & - & - & $0.57^{+0.05}_{-0.04}$ & $0.57^{+0.05}_{-0.04}$ \\
\enddata
\end{deluxetable*}

\startlongtable
\begin{deluxetable}{ll}
\tablecaption{FGS residuals extracted from \citet{fgs_eps_eri_benedict_2006} in the reported $\xi$ direction.\label{tab:fgs-eta}}
\tabletypesize{\scriptsize}
\tablehead{\colhead{epoch [MJD]} & \colhead{Position Residual $\eta$ [mas]}}
\startdata
  51946.5 & -1.84392 \\
  51946.5 & -1.81306 \\
  51946.8 & -1.31929 \\
  51947.0 & -1.01754 \\
  51980.5 & -1.72745 \\
  51980.5 & -1.69316 \\
  51981.2 & -2.54699 \\
  51981.3 & -2.46126 \\
  51981.3 & -2.39268 \\
  51981.4 & -2.31724 \\
  51981.4 & -2.25209 \\
  51981.7 & -1.84747 \\
  51982.0 & -1.27826 \\
  51982.0 & -1.25769 \\
  51982.1 & -1.23369 \\
  51982.1 & -1.1171 \\
  51982.2 & -1.00051 \\
  51982.2 & -0.973082 \\
  51982.3 & -0.856497 \\
  51983.2 & -1.45314 \\
  52179.3 & -2.25963 \\
  52179.3 & -2.22819 \\
  52179.3 & -2.17962 \\
  52179.4 & -2.10532 \\
  52179.4 & -2.08532 \\
  52179.7 & -1.54811 \\
  52179.9 & -1.20236 \\
  52179.9 & -2.84827 \\
  52180.0 & -2.81112 \\
  52180.0 & -2.77683 \\
  52180.0 & -2.40365 \\
  52180.0 & -2.72826 \\
  52180.5 & -1.9996 \\
  52180.5 & -1.95674 \\
  52180.5 & -1.87387 \\
  52180.6 & -1.80815 \\
  52180.6 & -1.77386 \\
  52180.8 & -1.43096 \\
  52181.1 & -1.01091 \\
  52181.8 & -1.48525 \\
  52308.3 & -3.63452 \\
  52308.4 & -3.57451 \\
  52308.5 & -3.29734 \\
  52308.6 & -3.16018 \\
  52308.6 & -3.1316 \\
  52308.7 & -2.98016 \\
  52308.8 & -2.92872 \\
  52308.8 & -2.85443 \\
  52309.0 & -2.62297 \\
  52309.1 & -2.48581 \\
  52309.1 & -2.43152 \\
  52309.1 & -2.41723 \\
  52309.2 & -2.18006 \\
  52309.3 & -2.11148 \\
  52309.3 & -2.07148 \\
  52309.7 & -1.4714 \\
  52309.8 & -2.90586 \\
  52309.9 & -1.12279 \\
  52310.0 & -0.977059 \\
  52310.1 & -0.854187 \\
  52310.3 & -0.559866 \\
  52311.3 & -0.51415 \\
  52497.0 & -3.48657 \\
  52497.3 & -3.02366 \\
  52498.0 & -3.63802 \\
  52498.3 & -3.20654 \\
  52498.4 & -3.01224 \\
  52498.5 & -2.78364 \\
  52499.5 & -2.92937 \\
  52499.8 & -2.37502 \\
  52503.6 & -3.07798 \\
  52503.7 & -2.86081 \\
  52503.7 & -2.82937 \\
  52504.0 & -2.45504 \\
  52505.0 & -2.44647 \\
  52505.4 & -3.54375 \\
  52506.3 & -2.06929 \\
  52508.5 & -3.62092 \\
  52509.5 & -1.955 \\
  52510.6 & -1.88357 \\
  52512.9 & -3.27803 \\
  52513.8 & -3.56664 \\
  52514.1 & -2.978 \\
  52514.3 & -2.64653 \\
  52535.5 & -3.98962 \\
  52535.7 & -3.68959 \\
  52535.7 & -3.65816 \\
  52536.2 & -2.90092 \\
  52537.5 & -2.4323 \\
  52539.7 & -4.01821 \\
  52539.8 & -3.82962 \\
  52540.2 & -3.16097 \\
  52540.7 & -2.3523 \\
  52542.4 & -3.06096 \\
  52543.7 & -2.58091 \\
  52544.0 & -2.20086 \\
  52642.1 & -2.79556 \\
  52642.9 & -3.17846 \\
  52642.9 & -3.1156 \\
  52643.8 & -1.664 \\
  52645.5 & -2.28979 \\
  52646.2 & -2.97845 \\
  52646.4 & -2.51268 \\
  52646.5 & -2.43553 \\
  52646.6 & -2.23836 \\
  52649.4 & -2.8413 \\
  52650.1 & -3.40137 \\
  52650.5 & -2.72129 \\
  52651.5 & -2.7613 \\
  52673.0 & -3.68148 \\
  52673.6 & -2.71851 \\
  52673.6 & -1.03259 \\
  52673.7 & -2.57278 \\
  52674.2 & -1.65267 \\
  52678.4 & -3.3929 \\
  52678.8 & -2.6871 \\
  52679.1 & -2.30419 \\
  52679.4 & -3.42719 \\
  52680.8 & -2.91285 \\
  52682.2 & -3.95012 \\
  52682.3 & -3.81868 \\
  52683.1 & -2.52423 \\
  52683.3 & -2.21277 \\
  52683.8 & -1.41553 \\
  52684.0 & -1.12121 \\
  52685.9 & -3.13575 \\
  52686.5 & -3.82726 \\
  52686.7 & -1.86702 \\
  52686.8 & -3.33006 \\
  52688.2 & -2.78428 \\
  52689.7 & -1.98133 \\
  52689.8 & -1.87846 \\
  52860.9 & -3.11348 \\
  52862.1 & -2.87346 \\
  52864.3 & -2.65916 \\
  52864.6 & -3.80215 \\
  52865.6 & -2.35055 \\
  52866.5 & -2.49057 \\
  52869.4 & -2.92206 \\
  52869.6 & -2.59917 \\
  52869.8 & -2.2477 \\
  52869.8 & -2.22198 \\
  52870.0 & -1.95623 \\
  52871.3 & -3.1621 \\
  52871.8 & -2.42201 \\
  52872.0 & -3.70788 \\
  52872.2 & -1.82194 \\
  52873.7 & -2.73634 \\
  52875.3 & -3.50215 \\
  52875.7 & -2.8135 \\
  52875.8 & -2.76206 \\
  53007.5 & -3.22257 \\
  53008.4 & -1.79097 \\
  53009.2 & -0.487956 \\
  53010.7 & -3.07971 \\
  53011.2 & -2.37105 \\
  53013.4 & -3.7998 \\
  53013.7 & -3.38833 \\
  53013.8 & -3.28831 \\
  53014.3 & -2.49108 \\\hline
\enddata
\end{deluxetable}

\startlongtable
\begin{deluxetable}{ll}
\tablecaption{FGS residuals extracted from \citet{fgs_eps_eri_benedict_2006} in the reported $\eta$ direction.\label{tab:fgs-xi}}
\tabletypesize{\scriptsize}
\tablehead{\colhead{epoch [MJD]} & \colhead{Position Residual $\xi$ [mas]}}
\startdata
  51946.5 & -1.84392 \\
  51946.5 & -1.81306 \\
  51946.8 & -1.31929 \\
  51947.0 & -1.01754 \\
  51980.5 & -1.72745 \\
  51980.5 & -1.69316 \\
  51981.2 & -2.54699 \\
  51981.3 & -2.46126 \\
  51981.3 & -2.39268 \\
  51981.4 & -2.31724 \\
  51981.4 & -2.25209 \\
  51981.7 & -1.84747 \\
  51982.0 & -1.27826 \\
  51982.0 & -1.25769 \\
  51982.1 & -1.23369 \\
  51982.1 & -1.1171 \\
  51982.2 & -1.00051 \\
  51982.2 & -0.973082 \\
  51982.3 & -0.856497 \\
  51983.2 & -1.45314 \\
  52179.3 & -2.25963 \\
  52179.3 & -2.22819 \\
  52179.3 & -2.17962 \\
  52179.4 & -2.10532 \\
  52179.4 & -2.08532 \\
  52179.7 & -1.54811 \\
  52179.9 & -1.20236 \\
  52179.9 & -2.84827 \\
  52180.0 & -2.81112 \\
  52180.0 & -2.77683 \\
  52180.0 & -2.40365 \\
  52180.0 & -2.72826 \\
  52180.5 & -1.9996 \\
  52180.5 & -1.95674 \\
  52180.5 & -1.87387 \\
  52180.6 & -1.80815 \\
  52180.6 & -1.77386 \\
  52180.8 & -1.43096 \\
  52181.1 & -1.01091 \\
  52181.8 & -1.48525 \\
  52308.3 & -3.63452 \\
  52308.4 & -3.57451 \\
  52308.5 & -3.29734 \\
  52308.6 & -3.16018 \\
  52308.6 & -3.1316 \\
  52308.7 & -2.98016 \\
  52308.8 & -2.92872 \\
  52308.8 & -2.85443 \\
  52309.0 & -2.62297 \\
  52309.1 & -2.48581 \\
  52309.1 & -2.43152 \\
  52309.1 & -2.41723 \\
  52309.2 & -2.18006 \\
  52309.3 & -2.11148 \\
  52309.3 & -2.07148 \\
  52309.7 & -1.4714 \\
  52309.8 & -2.90586 \\
  52309.9 & -1.12279 \\
  52310.0 & -0.977059 \\
  52310.1 & -0.854187 \\
  52310.3 & -0.559866 \\
  52311.3 & -0.51415 \\
  52497.0 & -3.48657 \\
  52497.3 & -3.02366 \\
  52498.0 & -3.63802 \\
  52498.3 & -3.20654 \\
  52498.4 & -3.01224 \\
  52498.5 & -2.78364 \\
  52499.5 & -2.92937 \\
  52499.8 & -2.37502 \\
  52503.6 & -3.07798 \\
  52503.7 & -2.86081 \\
  52503.7 & -2.82937 \\
  52504.0 & -2.45504 \\
  52505.0 & -2.44647 \\
  52505.4 & -3.54375 \\
  52506.3 & -2.06929 \\
  52508.5 & -3.62092 \\
  52509.5 & -1.955 \\
  52510.6 & -1.88357 \\
  52512.9 & -3.27803 \\
  52513.8 & -3.56664 \\
  52514.1 & -2.978 \\
  52514.3 & -2.64653 \\
  52535.5 & -3.98962 \\
  52535.7 & -3.68959 \\
  52535.7 & -3.65816 \\
  52536.2 & -2.90092 \\
  52537.5 & -2.4323 \\
  52539.7 & -4.01821 \\
  52539.8 & -3.82962 \\
  52540.2 & -3.16097 \\
  52540.7 & -2.3523 \\
  52542.4 & -3.06096 \\
  52543.7 & -2.58091 \\
  52544.0 & -2.20086 \\
  52642.1 & -2.79556 \\
  52642.9 & -3.17846 \\
  52642.9 & -3.1156 \\
  52643.8 & -1.664 \\
  52645.5 & -2.28979 \\
  52646.2 & -2.97845 \\
  52646.4 & -2.51268 \\
  52646.5 & -2.43553 \\
  52646.6 & -2.23836 \\
  52649.4 & -2.8413 \\
  52650.1 & -3.40137 \\
  52650.5 & -2.72129 \\
  52651.5 & -2.7613 \\
  52673.0 & -3.68148 \\
  52673.6 & -2.71851 \\
  52673.6 & -1.03259 \\
  52673.7 & -2.57278 \\
  52674.2 & -1.65267 \\
  52678.4 & -3.3929 \\
  52678.8 & -2.6871 \\
  52679.1 & -2.30419 \\
  52679.4 & -3.42719 \\
  52680.8 & -2.91285 \\
  52682.2 & -3.95012 \\
  52682.3 & -3.81868 \\
  52683.1 & -2.52423 \\
  52683.3 & -2.21277 \\
  52683.8 & -1.41553 \\
  52684.0 & -1.12121 \\
  52685.9 & -3.13575 \\
  52686.5 & -3.82726 \\
  52686.7 & -1.86702 \\
  52686.8 & -3.33006 \\
  52688.2 & -2.78428 \\
  52689.7 & -1.98133 \\
  52689.8 & -1.87846 \\
  52860.9 & -3.11348 \\
  52862.1 & -2.87346 \\
  52864.3 & -2.65916 \\
  52864.6 & -3.80215 \\
  52865.6 & -2.35055 \\
  52866.5 & -2.49057 \\
  52869.4 & -2.92206 \\
  52869.6 & -2.59917 \\
  52869.8 & -2.2477 \\
  52869.8 & -2.22198 \\
  52870.0 & -1.95623 \\
  52871.3 & -3.1621 \\
  52871.8 & -2.42201 \\
  52872.0 & -3.70788 \\
  52872.2 & -1.82194 \\
  52873.7 & -2.73634 \\
  52875.3 & -3.50215 \\
  52875.7 & -2.8135 \\
  52875.8 & -2.76206 \\
  53007.5 & -3.22257 \\
  53008.4 & -1.79097 \\
  53009.2 & -0.487956 \\
  53010.7 & -3.07971 \\
  53011.2 & -2.37105 \\
  53013.4 & -3.7998 \\
  53013.7 & -3.38833 \\
  53013.8 & -3.28831 \\
  53014.3 & -2.49108 \\\hline
\enddata
\end{deluxetable}

\end{document}